# Influence of the UV Environment on the Synthesis of Prebiotic Molecules

Sukrit Ranjan[1,2], Dimitar D. Sasselov[1]

*Harvard-Smithsonian Center for Astrophysics, Cambridge, MA, 02138, USA*

1. 60 Garden Street, Mail Stop 10, Cambridge, MA 02138, USA

2. sranjan@cfa.harvard.edu; 617-495-5676

# Abstract

Ultraviolet (UV) radiation is common to most planetary environments, and could play a key role in the chemistry of molecules relevant to abiogenesis (prebiotic chemistry). In this work, we explore the impact of UV light on prebiotic chemistry that might occur in liquid water on the surface of a planet with an atmosphere. We consider effects including atmospheric absorption, attenuation by water, and stellar variability to constrain the UV input as a function of wavelength. We conclude that the UV environment would be characterized by broadband input, and wavelengths below 204 nm and 168 nm would be shielded out by atmospheric $CO_2$ and water, respectively. We compare this broadband prebiotic UV input to the narrowband UV sources (e.g. mercury lamps) often used in laboratory studies of prebiotic chemistry, and explore the implications for the conclusions drawn from these experiments. We consider as case studies the ribonucleotide synthesis pathway of Powner et al (2009) and the sugar synthesis pathway of Ritson et al (2012). Irradiation by narrowband UV light from a mercury lamp formed an integral component of these studies: we quantitatively explore the impact of more realistic UV input on the conclusions that can be drawn from these experiments. Finally, we explore the constraints solar UV input places on the buildup of prebiotically important feedstock gasses like $CH_4$ and HCN. Our results demonstrate the importance of characterizing the wavelength dependence (action spectra) of prebiotic synthesis pathways to determine how pathways derived under laboratory irradiation conditions will function under planetary prebiotic conditions.

## 1. Introduction

Ultraviolet (UV) light plays an important role in the chemistry of prebiotic molecules. UV photons are energetic enough to affect the electronic structure of molecules by dissociating bonds, ionizing molecules, or exciting molecules into higher-energy states. These effects can degrade biologically important molecules, creating environmental stress and impeding abiogenesis (Sagan 1973; Cockell 2000). However, these same properties mean that UV light is an ideal candidate as a source of energy for Miller-Urey style synthesis of prebiotic molecules (Sagan and Khare 1971; Chyba and Sagan 1992; Pestunova et al. 2005). UV light has been invoked in prebiotic chemistry as diverse as the origin of chirality (Rosenberg et al. 2008), the synthesis of amino acids (Sarker et al. 2013), and the formation of ribonucleotides (Powner et al. 2009). Due to the greater fractional output of the young Sun in the UV compared to the modern Sun (Ribas et al. 2010; Claire et al. 2012), as well as the absence of biogenic UV-shielding $O_2$ and $O_3$ in the prebiotic terrestrial atmosphere, UV light is expected to have been a ubiquitous component of the prebiotic environment. Ferris and Chen (1975) estimate that for an ozone-free prebiotic atmosphere, UV light with $\lambda < 300$ nm contributed three orders of magnitude more energy than electrical discharges or shockwaves to the surface of the early Earth. UV light may have been the most abundant source of energy available for prebiotic chemistry.

Many experimental studies of prebiotic chemistry have sought to include the effects of UV irradiation. A large number of them are concerned with formation of prebiotic molecules on interstellar ices and cometary surfaces (see, e.g., Bernstein et al. 2000). Such studies usually use lamps or synchrotron sources in ultrahigh vacuum, and the UV output is below $\lambda < 160$ nm (see, e.g., Bernstein et al. 2002 and Öberg et al. 2009).

Prebiotic chemistry experiments in aqueous solution also often use UV lamps, because they are safe, stable and affordable UV sources. However, their output is often characterized by narrowband emission at specific wavelengths: for example, mercury lamps with primary emission at 254 nm are commonly used as proxies for prebiotic solar UV input (see, e.g., Moradpour and Kagan 1974; Ferris and Chen 1975; Kuzicheva and Gontareva 2001; Guillemin et al. 2004; Pestunova et al. 2005; Ferris et al. 2005; Powner et al. 2007; Simonov et al. 2007; Guzman and Martin 2008; Barks et al. 2010). However, solar UV input in that wavelength range is characterized by broadband emission. Many photoprocesses involving biological molecules are wavelength-dependent (e.g., Matsunaga et al. 1991); hence, conclusions drawn from simulations conducted using monochromatic UV light may not hold true under more realistic conditions. In addition, solar UV input also shapes atmospheric photochemistry, which may impact the availability of reactants for some of these prebiotic pathways, as well as the energy deposited at the surface.

In this work, we explore the impact of UV light on prebiotic chemistry in liquid water on planetary environments corresponding to the young Earth, and the implications for laboratory simulations. Therefore, we consider effects including atmospheric absorption, attenuation by water, and stellar variability, to estimate the UV input as a function of wavelength in prebiotically important environments. We compare these estimates to the output of UV lamps, and discuss the implications for laboratory studies like the ribonucleotide synthesis pathway of Powner et al. (2009) and the sugar synthesis pathway of Ritson and Sutherland (2012). We selected these two experiments as our case studies because they comprise the core part of a recently developed prebiotic chemistry network for the common origin of RNA, proteins, and lipid precursors (Patel et al. 2015). Irradiation by narrowband UV light from an Hg lamp formed an integral component of these studies: we quantitatively explore their viability under more realistic UV input. Finally, we determine the constraints solar UV input places on the buildup of prebiotically important feedstock gasses like $CH_4$ and HCN.

## 2. Background

### *2.1 UV Light and Prebiotic Chemistry*

In this subsection, we review the impact of UV light on prebiotic chemistry. By "prebiotic chemistry", we refer to the chemistry of small molecules, principally those containing C, H, N, O, P, or S, relevant to the origin of life. We focus on small molecules because large, complex molecules like proteins are not expected to be abundant prior to abiogenesis. We focus on molecules containing C, H, N, O, P, and S because those elements are the building blocks of life as we know it. We also include the chemistry of mineral catalysts that may be relevant to prebiotic chemistry. An example of a potentially relevant molecule is hydrogen cyanide, which may be a source of fixed nitrogen for organic molecules (Zahnle 1986). An example of a potentially relevant catalyst is the mineral montmorillonite ($(Na,Ca)_{0.33}(Al,Mg)_2(Si_4O_{10})(OH)_2 \cdot nH2O$), which has been shown to promote polymerization of nucleotides (Ferris et al. 1996).

UV light directly impacts the chemistry of small molecules. UV light can break molecular bonds (photolysis), produce secondary electrons (photoionization), and excite molecules out of the ground state (photoexcitation). All three of these mechanisms can affect prebiotic chemistry. We discuss the importance of each mechanism in the following.

Many evaluations of the role of UV light in prebiotic chemistry focus on photolysis of molecular bonds, stressing the potential of UV light to destroy populations of prebiotically interesting molecules. Photolyzed molecules can also recombine to form substances toxic to biological life today; for example, photolyzed water fragments can produce toxic hydrogen peroxide, $H_2O_2$ (Alizadeh and Sanche 2012). From this perspective, UV light is considered a biological stressor, and many disfavor high-UV environments from a habitability perspective (Sagan 1973; Cockell 2000). However, UV photolysis may also play a role in promoting prebiotic chemistry. For example, Sagan and Khare (1971) rely on UV photolysis of $H_2S$ to produce superthermal H atoms that collisionally provide activation energy for reactions involving hydrocarbons, such as dissociation of $CH_4$ for participation in subsequent reactions. Bond dissociation energies for prebiotically interesting molecules vary, but are often contained in the range 1-10 eV (23-230 kcal/mol), corresponding to photons of wavelengths $\lambda$=120-1200 nm. For example, the dissociation energy of the C=C bond in the alkene $C_2H_4$ is 7.55 eV (171 kcal/mol, $\lambda$=164 nm), and the bond dissociation energy of the H-OH bond in water is 5.15 eV (119 kcal/mol, $\lambda$=240 nm) (Blanksby and Ellison 2003).

Absorption of UV light can photoionize molecules, release free electrons (termed secondary electrons) into the surrounding medium. Photoionization of liquid $H_2O$ begins at low efficiencies at 6.0-6.5 eV and increases to 100% efficiency at 11.7 eV (270. kcal/mol, $\lambda$=106 nm) (Mozumder 2002; Bernas et al. 1997), while benzene, the simplest aromatic hydrocarbon, photoionizes at 9.3 eV (214 kcal/mol, $\lambda$=130 nm) at 350K (Lias and Ausloos 1978). Production of secondary electrons generates free radicals that can interact chemically with other compounds in the system. This can lead to detrimental effects for biological systems. For example, secondary electrons can induce strand breaks in DNA and damage proteins (Alizadeh and Sanche 2012). However, photoionization can also drive relevant prebiotic chemistry. For example, Ritson and Sutherland (2012) rely on photoionization of cyanocuprates to drive their synthesis pathway for simple sugars. Generation of secondary electrons by photoionization may also help explain the origin of chirality in biomolecules. Rosenberg et al. (2008) demonstrate generation of chiral excesses at the 10% level in butanol via selective bond cleavage by spin-polarized electrons from a substrate. They suggest spin-polarized electrons generated by UV light incident on magnetic substrates as a potential mechanism for the origin of chirality.

UV light can excite molecules (photoexcitation), promoting electrons from the ground state to a higher energy state. Excited molecules can decay to the ground state via fluorescence (emission of photon from excited state to ground) or phosphorescence (emission of photons over multiple transitions through lower-energy excited states on the way to ground). Excited molecules can also dissipate this energy vibrationally or collisionally. These pathways can transfer energy to other molecules, "sensitizing" them for further interactions. Finally, excited molecules can return to their ground states by rearranging their electronic orbitals (undergoing chemical changes), including breakage of the bond corresponding to the excited electron. An example of such a process is the hydration and deamination of the nucleobase cytosine into uracil (U). Under exposure to UV light, water is taken up by photoexcited cytosine to form a photohydrate. This photohydrate is unstable, and decays in part to uracil, releasing ammonia in the process (Peng and Shaw 1996). This is the mechanism for generation of uracil in the synthesis pathway discovered by Powner et al. (2009). Another example of photoexcitation-induced chemistry is the formation of thymine-thymine dimers in DNA (Matsunaga et al. 1991). Through such mechanisms, photoexcitation can influence prebiotic chemistry.

### *2.2. UV Light and the RNA World*

A major theory for the origin of life is the RNA world hypothesis (Gilbert 1986; McCollom 2013). Under this hypothesis, RNA was the original autocatalytic information-bearing molecule, with enzymatic functions being accomplished by short RNAs. Eventually, proteins supplanted RNA as enzymes due to greater diversity of monomers, and DNA supplanted RNA as an information-bearing molecule due to greater stability. This model is appealing because it resolves the metabolism/genetics-first debate by leveraging RNA's ability to fulfil both catalytic and structural and genetic roles, and it explains the origins of RNA's intermediary role in the modern "Central Dogma" of molecular biology. While other models for the origin of life do exist (see e.g., Yu et al. 2013), the RNA world remains a dominant hypothesis (Copley et al. 2007). Recent work, in particular that of Powner et al. (2009) and Ritson and Sutherland (2012), have furnished key steps toward a plausible prebiotic synthesis of RNA. In this subsection, we discuss the pathways discovered by these studies and their interaction with UV light.

#### *2.2.1. Powner et al. (2009) Pathway for Synthesis of Activated Pyrimidines*

A key challenge with the RNA world hypothesis is how the comparatively complex RNA molecule originated (McCollom 2013). Powner et al. (2009) achieved a remarkable step forward with their synthesis of the activated pyrimidine ribonucleotides cytosine and uracil under prebiotically plausible conditions. This pathway is the first plausible candidate for the synthesis of activated ribonucleotides, and potentially fills in a missing step in the road to abiogenic RNA[1].

A key ingredient of the Powner et al. (2009) pathway is irradiation with UV light. UV irradiation, by a lamp with primary emission at 254 nm, plays two key roles in this pathway. First, UV irradiation destroys a number of competing pyrimidine molecules generated by the synthesis pathway. Indeed, the exceptional photostability of the pyrimidine ribonucleotides might in part explain why these particular pyrimidines were selected by evolution for incorporation into an informational polymer. Second, UV light is required to photoactivate ribocytidine to enable a partial conversion to ribouridine via hydration and deamination.

### *2.2.2. Ritson and Sutherland (2012) Pathway for Synthesis of Simple Sugars*

Irradiation by UV light is also a necessary element of a companion synthesis mechanism to the Powner et al. (2009) pathway, the Ritson and Sutherland (2012) synthesis of the two- and three-carbon sugars glycolaldehyde and glyceraldehyde from hydrogen cyanide (HCN) and the one-carbon sugar formaldehyde. These sugars are required for the synthesis of the pentose sugar ribonucleotide backbone in the Powner et al. (2009) process. Previously, the mechanism generally invoked to explain the prebiotic formation of sugars was the formose reaction, whereby formaldehyde polymerizes to form longer sugars. However, the formose reaction is nondiscriminate, meaning that it produces not only glycolaldehyde and glyceraldehyde but also longer sugars as well as structural isomers of the sugars. Additionally, the polymerization tends to run away, generating longer and longer chains until the products form an insoluble tar, useless to prebiotic chemistry (McCollom 2013; Ritson and Sutherland 2012). By contrast, the Ritson and Sutherland (2012) pathway uses a much more selective Kiliani-Fischer synthesis that generates a small number of products, including glycolaldehyde and glyceraldehyde in solution, available for further chemistry.

[1] Ferris et al. (1996) demonstrate mechanisms by which activated nucleotides can polymerize to form nucleic acids.

Ritson and Sutherland (2012) suggest that their synthesis relies on production of solvated electrons and protons via photoionization of the photocatalytic transition metal cyanide complex tricyanocuprate (I) ($Cu(CN)_3^{2-}$). Cyanocuprates of this type can be generated from solvated Cu+ and $CN^-$ ions, for example via reaction of copper sulfides with cyanide solution (Patel et al. 2015). Under irradiation from a mercury lamp with primary emission at 254 nm, tricyanocuprate (I) photoionizes to tricyanocuprate (II), releasing a solvated electron on the way. This electron transfer triggers a cycle during which HCN is reduced, initiating the Kiliani-Fischer synthesis. Efforts to achieve this effect via pulse radiolysis (exposure to a beam of accelerated electrons) were unsuccessful due to generation of additional radicals, which lead to a proliferation of reaction products. This suggests UV irradiation is required for this pathway.

Banerjee et al. (2014) suggest an alternate mechanism underlying the photoredox synthesis of simple sugars discovered by Ritson and Sutherland (2012). They conduct theoretical calculations to suggest that instead of photoionizing electrons from tricyanocuprate (I) to reduce HCN, the UV input excites the transition metal complex from its ground $S_0$ state to its excited $S_1$ state. The $S_1$ state then decays to the triplet $T_1$ state via intersystem crossing. This state favorably binds HCN. The resulting molecular complex can then relax by dissociation to $HCN^-$ and tricyanocuprate (II), from which point the cycle can proceed. This mechanism is appealing because it avoids generating free radicals, which (due to their high reactivity) might impair the selectivity of the reaction process.

The Ritson and Sutherland (2012) and Banerjee et al. (2014) photoredox pathways differ significantly in their wavelength dependence on the input UV energy. The Ritson and Sutherland (2012) process simply calls for the photoionization of the tricyanocuprate (I) complex. This process should proceed with approximately equal quantum efficiency for input photons with energy exceeding the ionization energy of tricyanocuprate (I). If the ionization energy of tricyanocuprate (I) is $E_0$, then for $\lambda < \lambda_0 = hc/E_0$ the pathway should proceed; for $\lambda > \lambda_0$, it should not. By contrast, the Banerjee et al. (2014) pathway requires the excitation of tricyanocuprate from $S_0$ to $S_1$, which then decays to a reactive triplet state via intersystem crossing. Banerjee et al.

(2014) calculate this initial excitation occurs upon the absorption of photons with energies corresponding to a wavelength of 265 nm. Under the Banerjee et al. (2014) mechanism, the quantum efficiency of the process should peak at 265 nm.

## 3. Results

As discussed in Section 2, UV radiation has a key influence on prebiotic chemistry, and many experiments that seek to replicate prebiotic conditions have sought to incorporate it into their studies. However, many if not most of these studies have used narrowband UV lamps whose emission spectra are characterized by line emission. In this section, we provide estimates of the UV input in environmental conditions relevant to prebiotic chemistry and compare them to narrowband lamp input. We review factors including stellar output and activity, atmospheric attenuation, and aqueous shielding of UV flux. We also consider the constraints furnished by UV input on buildup of feedstock gases relevant to prebiotic chemistry. Our objective is to furnish guidance to experimentalists seeking to better simulate the prebiotic environment in laboratory settings.

### *3.1. Solar UV Output in the Prebiotic Era*

The dominant source of UV light in the prebiotic era, as today, was the Sun. The young Sun around the era of abiogenesis was as much as 30% dimmer and significantly more active than the Sun today. Studies of the star $\kappa^1$ Ceti, which is a proxy for the 3.7-4.1 Ga Sun, show that despite being less luminous than the modern Sun overall, $\kappa^1$ Ceti's emission exceeds solar emission for wavelengths shorter than 170 nm by 10−15% (Ribas et al. 2010). This is due to a higher level of magnetic activity powered by a more rapidly rotating stellar dynamo. Based on observations of this solar analog, we expect the young Sun to have delivered far more of its energy in the UV wavelengths than does the Sun today. As an initial estimate of the young Sun's luminosity, we turn to the model of Claire et al. (2012), which uses data from the Sun and solar analogs to calibrate wavelength and time-dependent scalings for the modern Sun's emission spectrum. The net data product is a model for the solar spectrum through time. For purposes of our study, we choose 3.9 Ga as the era of abiogenesis. This period coincides with the end of the Late Heavy Bombardment (LHB) and is consistent with available geological and fossil evidence for early life (see Appendix A for details).

#### *3.1.1. Effect of Stellar Variability*

Shortwave UV flux ($\lambda < 170$ nm) is formed in the upper solar atmosphere, which is composed of high-temperature plasma sensitive to phenomena such as stellar activity and flares (Ribas et al. 2010). As the young Sun formed a larger fraction of its emission from shortwave radiation (Cnossen et al. 2007), this argues that the young Sun should have had stronger fractional variations in UV output than the present day. What level of variability might we expect, and how might this variability affect biologically relevant molecules?

We can obtain an initial estimate of young Sun UV variability by using the modern Sun as a proxy for the young Sun. Since we expect the young Sun to have been more variable than the modern Sun, this estimate can be interpreted as a lower bound on the young Sun's variability. Thuillier et al. (2004) present composite spectra of the Sun synthesized from multiple data sources during the ATLAS 1 (March 1992) and ATLAS 3 (November 1993) space shuttle missions. These spectra correspond to moderately high (ATLAS 1) and low (ATLAS 3) periods of solar activity, and the epochs of observation "span half of the solar cycle amplitude in terms of the Mg II and F10.7 indices" (Thuillier et al. 2004). These data are accurate to 4% for the data spanning λ=122-400 nm. Figure 1a presents the ratio between these two measured spectra. UV variability is higher at shorter wavelengths due to greater relative emission from the hot outer atmosphere of the Sun. Line variability reaches as high as 20% at sampling of 0.05 nm over the temporal and wavelength range presented, but the highest-variability lines are also the short-wavelength ($\lambda < 200$ nm) lines most strongly screened by atmospheric absorbers like $CO_2$ and $H_2O$. For most of the unshielded $\lambda > 200$ nm region, variability is less than 5%. However, the Mg II k and h lines at 279.6 and 280.3 nm (Ayres and Linsky 1980) are variable at the 11% and 8% levels respectively, and these are not strongly shielded by expected atmospheric absorbers. Since the ATLAS 1 and 3 periods spanned only about half the amplitude of the solar cycle as measured by the Mg II index, over the full solar cycle we may expect variance at levels as high as 20% in these lines. We note these lines are comparatively narrow, with the entire Mg II line complex having a width of ~ 2 nm.

We obtain a second estimate of the UV variability of the young Sun by considering the effects of a flare on the stellar spectrum, as modeled by Cnossen et al. (2007). Cnossen et al. (2007) estimate the emission spectra of the young Sun in flaring and non-flaring states. To do so, they divide the emission of the 4-3.5 Ga Sun into components due to the photosphere and the outer-atmosphere (corona+chromosphere). Cnossen et al. (2007) estimate photospheric emission by scaling the emission of the modern Sun by 75%. To estimate the emission due to the outer atmosphere, Cnossen et al. (2007) use the young solar analog $\kappa^1$ Ceti. $\kappa^1$ Ceti is a $M_\star=1.04M_\odot$, [Fe/H]=0.10 ± 0.05, T=0.4 − 0.8 Gyr analog to the 3.7-4.1 Ga Sun (Ribas et al. 2010). Cnossen et al. (2007) use UV and X-ray observations of $\kappa^1$ Ceti to derive an emission model for the outer atmosphere of the star. Adding this outer-atmosphere component to the previously derived photospheric component yields an estimate of the UV spectrum of the 3.7-4.1 Ga non-flaring Sun. To estimate the UV spectrum of the flaring Sun, they scale a solar flare to $\kappa^1$ Ceti to form a new emission model for the outer atmosphere, and proceed as before. Figure 1b presents the Cnossen et al. (2007) ratio between the flaring and non-flaring young Sun models. For wavelengths shorter than 150 nm, chromospheric/coronal emission dominates, and there exist differences in flux that exceed a factor of 100. However, wavelengths shorter than 210 nm are shielded by atmospheric absorption, muting their impact on terrestrial prebiotic chemistry. For wavelengths longer than 210 nm, photospheric emission dominates, and variability is again low. The maximum UV variability at wavelengths longer than 200 nm is found in the 210-215 nm bin, which displays variability at the 14% level.

We sought to obtain a quantitative estimate of the impact of variability on prebiotic chemistry. We noted that the variable Mg II h and k lines are nearly coincident with absorption peaks of some ribonucleotides (see Figure 2). Absorption of UV photons by ribonucleotides can drive a variety of chemistry, ranging from photodegradation to photohydration/degradation. To gain a quantitative estimate of the potential photochemical impact of the variability of the Mg II h and k lines on the absorption of UV photons by ribonucleotides, we compute the difference in the rate of ribonucleotide photon absorption for the low and medium-high activity Thuillier et al. (2004) solar spectra. We weight these composite spectra with the absorption spectra corresponding to ribocytidine (at pH=7.9 and 2.5) and ribouridine (at pH=7.6 and 3.2), taken from Voet et al. (1963)[2]. We integrate these weighted spectra across the photochemically relevant 200-300 nm, and compare the resulting photoabsorption rates. Integrated across the $\lambda$=200-300 nm UV window, the variation in photoabsorption rate is $\leq 0.4\%$ for all ribonucleotides, with the variation ranging from 0.09% (ribocytidine, pH=2.5) to 0.4% (ribocytidine, pH=7.9). Note that our data cover only half of the solar cycle amplitude; hence, we can expect that over a full solar cycle, we would see twice the variation in photoabsorption rates. The change in ribonucleotide UV photoabsorption rate due to the solar cycle as measured from the modern sun is small (< 0.65%) in the broadband (λ=200-300 nm), though in narrowband (0.05 nm) it can change by as much as 11% (Mg II line).

---

[2] Note this procedure implicitly assumes an optically thin solution of ribonucleotides. At sufficiently high concentrations the solution will be optically thick and essentially all incident flux will be absorbed

We perform a similar analysis with the Cnossen et al. (2007) young sun flare models. We convolve their flaring and non-flaring models against the ribonucleotide absorption spectra from Voet et al. (1963). Integrated across λ=200-300 nm, the variations in photoabsorption rates are 0.2-0.3%, with the minimum and maximum variation again corresponding to ribocytidine at pH=2.5 and 7.9, respectively. As with solar cycle-driven variation in UV output, flare-driven variations in UV output can affect the ribonucleotide UV photoabsorption rate by up to 14% in a narrow band (5 nm), but integrated across λ=200-300 nm variation in photoabsorption rates falls to < 1%.

Overall, UV variability in the λ > 204 nm region of the emission spectrum of the young Sun that is relatively unshielded by atmospheric absorbers (especially $CO_2$) is low, usually at the level of a few percent or less. UV variability within individual features can vary at the level of tens of percent, but these features are generally narrow, muting their effect on the total UV power being received by the Earth.

### *3.2. Attenuation of UV Light By Aqueous Environments*

Prebiotic chemistry, both on Earth and in space (e.g. on meteorite progenitors, see Glavin et al. 2012) is thought to have occurred in an aqueous environment. In this section, we explore attenuation of incident UV flux by water.

For the absorption spectrum of water in the UV, we draw upon the data presented in Segelstein (1981), compiled in Querry et al. (1991), and made available online by Jonasz (2007). This work aggregated previous measurements of the absorption coefficient $\kappa(\lambda)$ of liquid water near standard conditions from 101 − 1010 nm, and confirmed their validity with an electron sum rule calculation. Figure 3 presents the UV spectrum of the 3.9 Ga Sun derived from the models of Claire et al. (2012) filtered through varying depths of water. Water absorption rises sharply at wavelengths shorter than 173 nm, with 1 μm of water enough to reduce stellar flux by a factor of 10 or more for $\lambda < 168$ nm. 1 μm of water is enough to extinguish transmission by 6 orders of magnitude at $\lambda$=160 nm, while 78 m of water is required to extinguish flux at wavelengths $\lambda$=300 nm by an equivalent amount. This high level of absorption is due to a broad, strong absorption band (peaking at 65 nm) corresponding to a superposition of transitions associated with the photoionization and photodissociation of water (Wozniak and Dera 2007). For example, incident photons with wavelength 180 nm and shorter have enough energy to promote the $1b_1 \rightarrow 4a_1$ orbital transition, dissociating water to OH+H; for more details, see Mota et al. (2005) and Wozniak and Dera (2007). The implication is that even a thin film of water on an asteroid is enough to shield mineral chemistry from XUV flux ($\lambda < 168$ nm), but midrange UV flux ($\lambda > 168$ nm) can readily penetrate. For comparison, the fatty acid vesicles studied as model systems for protocell membranes in work like Mansy et al. (2008) and Hanczyc et al. (2003) are of order 100 nm (0.1 μm) in size.

The results described above are for pure water. Prebiotic synthesis requires solutes as input for chemical reactions. We use modern seawater as a guide to approximate how dissolved constituents might affect the absorption of a water layer. 96-97% of seawater molecules are $H_2O$. The remaining 3-4% of molecules are dominated by dissolved salts, but also suspended organic and inorganic particulates and dissolved bioresidue ("Gelbstoff") (Wozniak and Dera 2007; Jonasz and Fournier 2007). The addition of salts causes increased absorption in the far-ultraviolet, as well as enhanced scattering due to variations in refractive index due to salt concentration variations (Jonasz and Fournier 2007; Cleaves and Miller 1998). Bioresidue ("Gelbstoff") also absorbs in the UV, but in pre-abiogenesis waters it would not be present. Scattering due to suspended particulates is also an important effect in seawater, with particle sizes ranging from roughly 0.01-1000 μm (Jonasz and Fournier 2007). In general, inorganic suspensions are weak absorbers in the UV (Wozniak and Dera 2007). In summary, inorganic dissolved and suspended minerals have the effect of enhancing UV attenuation; hence the transmitted UV flux presented here should be interpreted as an upper bound to prebiotic conditions.

### *3.3. Attenuation of UV Light by the Terrestrial Atmosphere*

Estimating the UV input on the terrestrial surface requires computing the attenuation of UV flux due to the atmosphere. As Earth in the prebiotic era would lack the biogenic oxygen and ozone that play a dominant role in UV attenuation in modern Earth's atmosphere, we are unable to use the modern Earth's atmosphere as a proxy for the young Earth's, and must instead estimate an atmospheric model from available information.

#### *3.3.1. Constraints on the Prebiotic Atmosphere*

Relatively little is known about the Earth's atmosphere in the prebiotic (~ 3.9 Ga) era. Synthesizing the available constraints (see Appendix B), we find that what evidence we have points to an $N_2$-$CO_2$ dominated atmosphere, with a high enough concentration of greenhouse gases (e.g. $CO_2$, $CH_4$) to sustain liquid surface water. Volcanogenic gases such as $SO_2$ may also have been important constituents during periods of high volcanic activity, and for a warm planet water vapor would also be an important atmospheric constituent. We therefore focus on these four gases as the major absorbers to consider when estimating UV flux to the planetary surface.

We note that Wolf and Toon (2010) have suggested the possibility of hydrocarbon hazes in providing planetary greenhouse warming, similar to what is seen on Titan today. Such hazes could act as UV shields. However, the formation of such hazes requires high $CH_4$ production rates, corresponding to a $CH_4/CO_2$ abundance ratio of 0.1 (DeWitt et al. 2009). Given the absence of biogenic production and lack of significant volcanic production of $CH_4$ based on the redox state of the mantle, such $CH_4$ production rates are unlikely. We consequently do not focus on hydrocarbon hazes in estimating atmospheric attenuation of UV light.

*3.3.2. Attenuation Due to The Prebiotic Atmosphere*

To estimate atmospheric attenuation due to the prebiotic atmosphere, we use the atmospheric model of Rugheimer et al. (2015). This model uses a coupled radiative-convective model that includes the effects of climate, photochemistry, and radiative transfer. This model uses stellar input equivalent to the Sun at an age of 3.9 Ga as modeled by the Sun Through Time project (Claire et al. 2012), squarely within the 3.5-4.3 Ga age range of plausible abiogenesis. It assumes an overall atmospheric pressure of 1 bar and atmospheric mixing ratios of 0.9, 0.1, and $1.65\times10^{-6}$ for $N_2$, $CO_2$, and $CH_4$, respectively. It also assumes modern abiotic outgassing rates of gases such as $SO_2$ and $H_2S$, and includes humidity (water vapor). Hence, the model includes the four key absorbers identified in the previous section. The nitrogen partial pressures adopted in the model are consistent with the constraints measured by Marty et al. (2013) for the 3.5 Ga Earth. The high $CO_2$ abundance relative to the present day is consistent with an atmosphere dominated by volcanic outgassing. The trace methane level is adopted from Kaltenegger et al. (2007). When iterated to convergence, this model indicates a surface temperature above freezing, indicating consistency with the constraints from zircon that suggests at least transient surface liquid water on Earth during this era.

Figure 4 presents the attenuation of the UV flux of the 3.9 Ga sun due to the primordial atmospheric model of Rugheimer et al. (2015). Compared to the present day, the prebiotic Earth would have received much more > 204 nm radiation due to the absence of UV-shielding oxygen. The planetary surface would still have been shielded from extremely shortwave radiation due to atmospheric $CO_2$: this fiducial atmosphere reduces incident stellar flux by a factor of 10 or more for wavelengths shorter than 204 nm. These results are consistent with the prebiotic atmosphere study of Cnossen et al. (2007). High surface UV is also consistent with the findings of Farquhar et al. (2001) that the primitive Earth was exposed to high UV.

Figure 4 also presents the attenuation of the surficial solar flux due to a 1 m water layer. For surface water layers of thickness ≥ 1 m, UV light with wavelengths up to 224 nm are additionally shielded out, meaning that for water layers with depth ≥ 1 m, UV light at wavelengths less than 224 nm is not relevant to prebiotic chemistry. For water column depths < 1 m, attenuation of UV at wavelengths shorter than 200 nm is dominated by atmospheric absorption. At the 254 nm wavelength range probed by mercury lamps, the attenuation due to this fiducial prebiotic atmosphere is equivalent to that provided by 127 cm of liquid water. Hence water absorption dominates the UV environment at 254 nm at depths greater than 127 cm (e.g. in the deep ocean).

This atmospheric model assumes modern levels of volcanogenic input of $SO_2$ and $H_2S$. It is plausible that volcanism levels, and hence $SO_2$ and $H_2S$ abundances, were at least transiently higher on the young Earth (see e.g. Kaltenegger and Sasselov 2010). Higher $SO_2$ and $H_2S$ levels could significantly impact the surface UV environment because these gases are better UV shields than $CO_2$ (see Figure 5).

We are not aware of existing empirical constraints on $SO_2$ and $H_2S$ levels at 3.9 Ga. We can place a theoretical upper limit on the abundance of $SO_2$ by the work of Halevy et al. (2007), who explored potential sulfur cycle chemistry on Mars. Their work suggested that at $SO_2$ levels of $10^{-6} - 10^{-4}$, $SO_2$ starts supplanting $CO_2$ as the controlling agent for temperature, precipitation, weathering, and aquatic reservoir pH chemistry, i.e. a sulfur cycle starts dominating over the carbon cycle. We take $10^{-4}$ as the upper bound on $SO_2$ level at 3.9 Ga. We constrain $H_2S$ levels by assuming the $[H_2S]/[SO_2]$ outgassing ratio to be the same as the present day value. Halmer et al. (2002) find the annual volcanic $SO_2$ and $H_2S$ fluxes to the atmosphere to be $15 - 21\times10^{12}$ g and $1.5 - 37.1\times10^{12}$ g, respectively, corresponding to an $[H_2S]/[SO_2]$ ratio of .29-9.9. We therefore adopt an upper bound of $10^{-3}$ on $H_2S$ levels at 3.9 Ga.

For $\lambda$=200-300 nm, $\sigma_{SO2} /\sigma_{CO2}=3\times10^6$. At an $SO_2$ levels of $10^{-4}$, $[SO_2]/[CO_2]= 10^{-3}$. At this $SO_2$ level, attenuation from $SO_2$ would outpace extinction from $CO_2$ by 3 orders of magnitude. Similarly, integrated from 200-300 nm, $\sigma_{H2S}/\sigma_{CO2}=2\times10^6$. At an $H_2S$ level of $10^{-3}$, this corresponds to extinction from $H_2S$ outpacing extinction from $CO_2$ by 4 orders of magnitude. It is difficult to confidently describe the impact of epochs of high volcanism on the UV surface environment, given the uncertainties in $SO_2/H_2S$ levels as well as secondary photochemical effects, e.g. formation of UV-shielding hazes Tian et al. (2010); Wolf and Toon (2010). Further modelling, using a framework capable of accounting for high $SO_2$ cases, is required. However, it seems plausible that episodes of high volcanism may be low-UV epochs in the young Earth's history.

### *3.4. CO2 Shielding of Prebiotic Feedstock Gases From Photolysis*

Theoretical and experimental explorations of prebiotic chemistry often assume the availability of feedstock gases, substances that may plausibly have been present in the prebiotic era which furnish biologically useful forms of elements relevant to prebiotic chemistry. Examples include $CH_4$ as a source of reduced carbon, and HCN as a source of reduced nitrogen (Zahnle 1986).

One fundamental constraint on the abundance of such substances is photolysis. UV photolysis can destroy these compounds as they are formed, preventing them from being available to participate in prebiotic chemistry. UV attenuation from high $CO_2$ atmospheres, like the 0.1 bar $CO_2$ atmosphere invoked in Rugheimer et al. (2015), may be capable of shielding these gases from photolysis. However, such high $CO_2$ atmospheric models are ad hoc models: the $CO_2$ level is assumed. Effects such as the carbon cycle may present geophysical sinks not considered in these models that may make high $CO_2$ levels impossible to sustain. We adopt such ad hoc models because they are the best guess we have at conditions at 3.9 Ga. However, from the point of view of understanding what feedstock gases are plausibly accessible for prebiotic chemistry, it behooves us to consider how sensitive the abundance of some of these gases is to the $CO_2$ level.

In this section, we estimate the levels of $CO_2$ required to quench direct photolysis of the prebiotic feedstock gases methane and HCN. These molecules are often included in prebiotic chemistry studies, and photolysis is a major sink for both. We conduct this study by estimating the abundance of these gases under equilibrium conditions using a simple 1 source/1 sink model, and computing the $CO_2$ column density required to attenuate incoming UV flux to the point at which their abundances can build up to varying levels.

We assume the source of methane/HCN to be abiotic geochemical fluxes, and we assume the sink to be UV photolysis. To gain traction on the problem, we assume an isothermal atmosphere of ideal gases in hydrostatic equilibrium. Following Rugheimer et al. (2015), we take our young Earth atmosphere to be $N_2/CO_2$ dominated with a surface pressure of 1 bar. We assume the mixing ratio of the feedstock gas under consideration to remain constant until a height $z_0$, whereupon it goes to zero. Lastly, we assume the feedstock gas population to be optically thin, i.e. we ignore self-shielding. This means our estimates will be upper bounds on the amount of $CO_2$ required to permit a given methane/HCN level. We take our top-of- atmosphere input $\varphi_0$ to be the UV spectrum of the 3.9 Ga sun computed from the models of Claire et al. (2012), spanning a wavelength range of 1-300 nm at 0.1 nm resolution. We obtain our $CO_2$ cross-sections $\sigma_{CO2}(\lambda)$ from Huestis and Berkowitz (2010), who compile $CO_2$ cross-sections at 300 K from 0.12-201.6 nm. Appendix C presents the detailed calculations used to compute these estimates. The code used to implement these calculations is available at https://github.com/sukritranjan/RanjanSasselov2015.

We emphasize that these calculations are not intended as realistic models of the early Earth's atmosphere. Rather, they are intended to estimate the column density of $CO_2$, $N_{CO2}$, that is required to quench photolysis to the degree required to build up the surficial partial pressure of a given feedstock gas to a level $P_{gas}$. We emphasize that achieving this $CO_2$ level does not automatically imply that the gas will build up to this degree; rather, if $CO_2$ levels build up to or past $N_{CO2}$, the constraint of photolysis on feedstock gas buildup to $P_{gas}$ is removed, though other constraints might remain.

### *3.4.1. $CO_2$ shielding of $CH_4$ from Photolysis*

Methane is important to prebiotic chemistry. Methane constitutes a reservoir of reduced carbon potentially useful to the synthesis of prebiotically interesting molecules (see e.g. Zahnle 1986, Ferris and Chen 1975). Indeed, early prebiotic chemistry studies, including the Miller-Urey experiment, were conducted in reduced gas mixes with methane as a major constituent (Kasting and Brown 1998). $CH_4$ levels indirectly control the local oxidation state, affecting the prebiotic chemistry that can proceed. The buildup of methane is limited by photolytic processes, including direct photolysis as well as interaction with O and OH radicals (whose production is also dominated by photolysis) (Rugheimer et al. 2015). $CO_2$ shielding can protect methane lower in the atmosphere from photolysis.

We estimate the column density of $CO_2$, $N_{CO2}$, required to shield $CH_4$ buildup to a level $P_{CH4}$ by the formalism presented in Appendix C. We choose $z_0$=50 km because the model of Rugheimer et al. (2015) shows a fall-off in $CH_4$ mixing ratio at this altitude. We take the rate of supply of methane to the atmosphere, S, to be equal to the present-day abiotic atmospheric flux of methane. Emmanuel and Ague (2007) estimate the present-day abiotic flux of methane to the atmosphere from serpentinization at mid-ocean ridges, volcanic emission, and other geothermal sources to be S~2.3 Mt/y=$7.3\times10^4$ g/s=$2.7\times10^{27}$ $s^{-1}$. For comparison, for a 1 $M_\oplus$, 1 $R_\oplus$ planet, Guzmán-Marmolejo et al. (2013) estimate a maximum abiotic methane production rate of $9.2\times10^4$g/s, which is consistent with the Emmanuel and Ague (2007) estimate to 26%. We adopt the Emmanuel and Ague (2007) estimate for S. We obtain our methane cross-sections from Au et al. (1993), who provide cross-sections from 5.6-165 nm for 298 K methane. The methane cross-section data set the limits of the wavelength range considered in this calculation; i.e., we calculated photolysis due to absorptions in the range of 5.6-165 nm.

We evaluate $P_{CH4}(z=0)$ for various values of $N_{CO2}$, holding all other parameters constant. We find that $CO_2$ column densities of $1.04\times10^{20}$ $cm^{-2}$, $2.59\times10^{20}$ $cm^{-2}$, and $4.87\times10^{20}$ $cm^{-2}$ are required to remove photolytic constraints on methane surface pressures of $10^{-9}$ bar, $10^{-6}$ bar, and $10^{-3}$ bar, respectively. Under our assumption of an isothermal atmosphere in hydrostatic equilibrium, these column densities correspond via the formula $P_{CO2}=N_{CO2}\times kT/H$ to shielding $CO_2$ partial pressures at $z=z_0=50$ km of $5.03\times10^{-6}$ bar, $1.25\times10^{-5}$ bar, and $2.35\times10^{-5}$ bar, respectively. Note that this calculation implicitly assumes all attenuation of the solar signal occurs for $z > z_0$; hence the $CO_2$ surface pressure requirements estimated here are upper bounds. For $N_{CO2}(z=50\text{km}) \geq 7.46\times10^{20}$ cm$^{-2}$ ($P_{CO2}(z=50\text{km}) > 3.60\times10^{-5}$ bar), $P_{CH4}(z=0) \geq 1$ bar. As this exceeds the total pressure of the atmosphere, this indicates that past this column density of $CO_2$, photolysis is no longer a constraint on $CH_4$ abundance. $P_{CO2}=3.60\times10^{-5}$ bar at $z_0=50$ km of altitude corresponds to a surface partial pressure of $CO_2$ of $1.49\times10^{-2}$ bar. Hence, a surface partial pressure of $CO_2$ of 0.015 bar is required to shield $CH_4$ from photolysis up to an altitude of $z_0=50$ km. On the other hand, if we are willing to restrict $CH_4$ to the bounds of the modern troposphere ($z_0 \approx 17$ km), then a $CO_2$ surface partial pressure of only $2.77\times10^{-4}$ bar is required to remove photolytic constraints on $CO_2$ buildup. If we further restrict $CH_4$ to the bottom 1 km of the atmosphere, then a $CO_2$ surface partial pressure of $3.60\times10^{-5}$ bar will suffice to shield the $CH_4$. Table 1 summarizes these findings. At even $10^{-4}$ bar levels, $CO_2$ is able to shield $CH_4$ in the troposphere from photolysis. Hence, under our assumptions, we may expect photolysis to not constrain the availability of $CH_4$ in chemistry in the lower atmosphere for a wide range of $CO_2$ surface pressures.

*3.4.2. $CO_2$ shielding of HCN from Photolysis*

Like methane, HCN is a key feedstock gas for prebiotic chemistry. HCN provides a biologically accessible source of reduced nitrogen (Zahnle 1986). Its presence has been invoked for a variety of prebiotic chemistry studies (see e.g., Ritson and Sutherland 2012, Ferris and Hagan 1984, Orgel 2004).

Following Section 3.4.1, we estimate the column density of $CO_2$, $N_{CO2}$, required to shield HCN buildup to a level $P_{HCN}$ by the formalism presented in Appendix C. Following Zahnle (1986), we take the rate of supply of HCN to the atmosphere, S, to be equal to 0.1× the methane flux. We adopt the same value for the methane flux as in Section 3.4.1, so $S \sim 2.7\times10^{26}$ $s^{-1}$. We obtain our HCN cross-sections from Nuth and Glicker (1982), who provide cross-sections from 100.5-299.5 nm. Based on Lee (1980), we assume that all absorptions lead to photolysis.

Unlike methane, which in the UV absorbs only at wavelengths shorter than 165 nm (Romanzin et al. 2005), HCN absorbs until 190 nm. Due to higher solar output and reduced $CO_2$ shielding at these longer wavelengths, HCN absorbs orders of magnitude more photons than $CH_4$ and suffers a concomitant increase in photolysis rates. As a result, much higher levels of $CO_2$ are required to shield equivalent amounts of HCN. Indeed, while the column density of $CO_2$ required to shield $CH_4$ from photolysis up to an altitude of $z_0$=50 km corresponds to a $CO_2$ surface pressure of 0.015 bar, the amount of $CO_2$ shielding required to remove photolytic constraints on HCN at this altitude would correspond to 4.75 bars of $CO_2$ at the surface – far exceeding the 0.1 bar of surface $CO_2$ assumed in our ad hoc model atmosphere.

However, if we restrict HCN to the bounds of the modern troposphere ($z_0 \approx 17$ km), absorption corresponding to a much lower surface pressure of $CO_2$ is required to shield HCN from photolysis. At $z_0$=17 km, $CO_2$ column densities of $2.30\times10^{20}$ cm$^{-2}$, 4.78×1022 cm$^{-2}$, and $1.37\times10^{23}$ cm$^{-2}$ are required to removed photolytic constraints on HCN surface pressures of $10^{-9}$ bar, $10^{-6}$ bar, and $10^{-3}$ bar, respectively. For $N_{CO2} > 2.35\times10^{23}$ cm$^{-2}$, $P_{HCN} > 1$ bar, indicating the photolytic constraint on HCN has been lifted. This corresponds to $P_{CO2}$(z=17 km)=$1.11\times10^{-2}$ bar, and $P_{CO2}$(z=0)=$8.81\times10^{-2}$. Table 2 summarizes the level of $CO_2$ required to remove the constraint of photolysis on HCN up until different heights $z_0$. Overall, while very high (> 1 bar) levels of $CO_2$ are required to shield HCN from photolysis at high altitudes ($z_{0=}$50 km), $CO_2$ levels of ≈ 0.1 bar are adequate to remove photolytic constraints on HCN in the troposphere (up to $z_0$=17 km), and levels of 0.01 bar can shield HCN below $z_0$=1 km. Compared to $CH_4$, much higher levels of CO2 are required to shield HCN from photolysis, and photolysis constrains HCN to be lower in the atmosphere than $CH_4$. Under our assumptions, surface $CO_2$ levels corresponding to our ad-hoc prebiotic atmospheric model (0.1 bar) are adequate to shield HCN in the troposphere from photolysis.

## 4. Discussion

### *4.1. Lessons for Laboratory Simulations*

The prebiotic UV environment, both in space and on the terrestrial surface, was likely characterized by broadband UV exposure. On the ground, the terrestrial atmosphere will shield UV radiation shortward of 204 nm. However, there remains high UV throughput at wavelengths longer than 204 nm, even with additional shielding from layers of water as much as a meter thick. Sites of relevant prebiotic chemistry in space will lack this atmospheric shielding. However, even micron-thick layers of water will shield out UV radiation shortward of 168 nm.

This yields a few immediate lessons for prebiotic chemistry simulations involving UV light. For simulations of prebiotic chemistry on Earth, wavelengths shorter than 204 nm are not accessible. Lamps that operate at such wavelengths should therefore not be used as UV sources in prebiotic simulations. For example, ArF eximer lasers (193 nm) (e.g., Pestunova et al. 2005), and Hg lamps with strong emission at the 184.9 nm line (e.g., Ferris and Chen 1975) are disfavored for use in simulations of surficial or near-surficial prebiotic chemistry. Similarly, for studies of aqueous chemistry on bodies not shielded by an atmosphere, lamps with emission shortward of 168 nm are disfavored.

More generally, energy is delivered to the prebiotic environment along a broad UV band. Experiments that use narrow-band lamps risk under-activating chemical pathways with photoactivation curves that are not coincident with the lamp emission peaks, while over-activating pathways whose photoactivation curves are coincident with lamp emission. As such, broadband UV sources such as Xenon arc-discharge lamps should be favored over narrowband sources in prebiotic chemistry. Such broadband sources should be fitted with filters to remove $\lambda < 204$ nm and $\lambda < 168$ nm radiation, as appropriate.

Estimates of UV variability derived from the modern Sun and from flare models of $\kappa^1$ Ceti suggest that narrowband variability is low, generally on the order of a few percent or less and never more than ~ 20%. Integrated from 200-300 nm, this corresponds to a < 1% variance in the number of photons absorbed by biomolecules such as ribonucleotides. Based on these data, we argue that stellar UV variability is a less important phenomenon to consider when designing laboratory studies of prebiotic chemistry.

In addition to the differences in the shape of the input UV spectrum, UV lamps provide higher intensity radiation than would have been accessible at the planetary surface. For example, the lamp apparatus used in Powner et al. (2007) and Powner et al. (2009) can be expected to deliver $6\times10^{15}$ photons/s/cm$^2$ to the sample integrated across its 254 nm emission feature at a distance of 1.9 cm. In comparison, we expect the surficial solar input to have delivered $4\times10^{13}$ photons/s/cm$^2$ from 250-260 nm, and $1\times10^{15}$ photons/s/cm$^2$ from 200-300 nm. Hence we expect the lamp to deliver 1-2 orders of magnitude more flux to the sample than the natural environment. This difference has in the past been dismissed under the argument that increasing the UV flux serves simply to accelerate the UV photochemistry to timescales more readily accessible in the laboratory. However, care must be taken in the interpretation of such studies to ensure that nonlinearities at lower fluence levels (e.g., due to backreactions) do not disrupt the pathway. One way to probe such effects is to measure reaction rate as a function of fluence level and verify linearity down to natural fluence levels.

### *4.2. Narrowband vs. Broadband Input*

We have demonstrated that the natural prebiotic environment was characterized by broadband UV input, as compared to the narrowband input provided by sources like mercury lamps. However, the question remains as to the impact of using narrowband lamps instead of more realistic natural input: are narrowband sources viable proxies for prebiotic UV flux? We explore this question through the case studies of the Powner et al. (2009) and Ritson and Sutherland (2012) pathways.

#### *4.2.1. Implications for Powner et al. (2009) Process*

The Powner et al. (2009) pathway for synthesis of activated pyrimidine ribonucleotides was derived under narrowband 254 nm emission from a mercury lamp. In this section, we explore whether this pathway can proceed under our modeled prebiotic UV input.

UV light impacts the Powner et al. (2009) pathway in two ways. One is a relatively straightforward photohydration and deamination of cytosine, which is the mechanism for production of uracil. The second way is by destroying competitor pyrimidine nucleosides and nucleotides generated by the phosporylation reaction. Non-biogenic pyrimidine molecules generated along with the biogenic ribonucleotides photolyzed at higher rates than the biogenic molecules, amplifying the population of biogenic molecules over time. However, this phenomenon was observed only at 254 nm. It is unknown whether at different wavelengths competitor molecules might have higher survivability than the ribonucleotides, negating the amplification mechanism used in this pathway. It has been argued that the biogenic nucleobases emerged as informational polymers because of exceptional stability to UV radiation (see, e.g., Mulkidjanian et al. 2003). Under this hypothesis, ribocytidine and ribouridine are more photostable than competitor molecules across all wavelengths, in which case 254 nm radiation would be a good proxy for prebiotic UV input. However, this hypothesis has not been proven. It remains possible that at other wavelengths, other competitor molecules might emerge as more stable than ribouridine and ribocytidine. In this case, 254 nm radiation would be a poor proxy for prebiotic UV input. Empirical measurements of nucleotide photostability as a function of wavelength are required to differentiate between these possibilities.

We illustrate this quantitatively with a numerical experiment. Powner et al. (2009) found irradiation to enhance the population of the biogenic β form of the molecule relative to the α form also produced previously in the synthesis pathway. This implies that the photolysis rate of the β form should be less than that of the α form. We compute photolysis rates for the α (non-biogenic) and β (biogenic) stereoisomers of ribocytidine as a function of wavelength under irradiation by 1) a Pen-Ray mercury lamp of the type used by Powner et al. (2009) and 2) under modeled natural prebiotic input. To do so, we must define action spectra for photolysis of each of these molecules. These action spectra can be computed as the product of absorption spectra (fraction of incident photons absorbed) and quantum efficiency functions (QEF, fraction of absorbed photons leading to photolysis). We assume that both forms of the ribocytidine molecule share the same absorption spectrum since they share similar chromophores, which we represent by the absorption spectrum of pH=7.9 ribocytidine measured by Voet et al. (1963). For comparison, the Powner et al. (2009) experiments were conducted at pH=6.5. As the QEF for photolysis of α and β are not available, we assume functional forms for them. We represent the QEF of β ribocytidine photolysis by a flat line at 0.4, and the QEF of α ribocytidine photolysis by a step function with value 0.6 between 250 and 260 nm, and 0.2 elsewhere. We emphasize that these QEFs are not physically motivated, and should not be taken to be representative of the true QEFs. Rather, they are constructed to illustrate the point that QEFs may exist that permit β ribocytidine to be more stable than the α form at wavelengths accessed by the mercury lamp, but not at other wavelengths.

Figure 6 presents the formation of the phototolysis rate calculation. We represent the prebiotic flux with the emergent surface spectrum computed by Rugheimer et al. (2015), corresponding to the radiation environment on the surface following attenuation by the atmosphere. We represent the lamp flux by the emission spectrum of a Pen-Ray 90-0012-01 UVP mercury lamp emitting $4.4\times10^4$ erg/s/cm$^2$ at a distance of 1.9 cm integrated across the 254 nm line. We compute the fraction of incident flux absorbed as a function of wavelength via the molar absorptivities from Voet et al. (1963); we assume a path length of 0.53 cm and a nucleotide concentration of 20 mM, chosen to accord with the experimental setup of Powner et al. (2009). At this concentration and path length, the absorption spectrum is optically thick across most of the wavelength range under consideration, implying the photolysis rate should depend only weakly on the spectral absorbance. We consider fluxes from 200-303 nm, chosen to encompass the 200-300 nm region of photochemical interest while avoiding truncating the wavelength bins of our computed prebiotic UV spectrum.

We convolve the two different incident flux distributions against the absorption spectrum and the assumed quantum efficiency curves to derive photoreaction rates as a function of wavelength. Figure 6 shows each of these input curves and the product of their convolution. Integrating over wavelength, we find that under these assumed QEFs, α-ribocytidine has a 47% higher photolysis rate under irradiation by lamp flux than β-ribocytidine, yielding the stability advantage to the biogenic β stereoisomer observed in the Powner et al. (2009) experiment. However, under irradiation by prebiotic UV input, α ribocytidine photolyzes at a rate 46% lower than β ribocytidine, meaning that under prebiotic input, the nonbiogenic stereoisomer would be amplified relative to the biogenic. This numerical experiment demonstrates that there exist QEFs consistent with the Powner et al. (2009) laboratory experiments that would nonetheless cause the reaction pathway to fail under exposure to a more realistic prebiotic UV environment. These results highlight the importance of employing realistic simulations of UV environments when conducting prebiotic chemistry studies. Additionally, it is important to characterize the wavelength dependence of the action spectra of the underlying photoprocesses of these pathways, e.g. by repeating the experiment at different irradiation wavelengths using tunable UV sources and measuring yield. Measuring these action spectra permits characterization of the underlying photochemistry, enabling extrapolation from laboratory results to prebiotic settings.

*4.2.2. Implications for Ritson and Sutherland (2012) Process*

The Ritson and Sutherland (2012) process also relies on narrowband mercury lamp emission at 254 nm. In this section, we explore how irradiation under our modeled natural prebiotic UV input would affect this process.

There are two presently postulated mechanisms by which UV light reduces HCN to drive the Ritson and Sutherland (2012) pathway. The first is the hypothesis outlined by Ritson and Sutherland (2012) whereby UV light photoionizes an electron from tricyanocuprate (I). In this case, any incident radiation at wavelengths shorter than a critical wavelength $\lambda_c$ will drive this synthesis, where $\lambda_c$ corresponds to a photon with energy equal to the work function of tricyanocuprate (I). Empirically, we know $\lambda_c \geq 254$ nm, since the reaction proceeded under action of 254 nm radiation. Alternately, Banerjee et al. (2014) suggest that the pathway is driven by UV excitement of tricyanocuprate (I) from the S0 to the S1 state. The energy difference between these states is calculated to correspond to 265 nm; consequently, 265 nm radiation should be much more efficient than 254 nm radiation at promoting this transition, and the Ritson and Sutherland (2012) experiments may underestimate the reaction rate. As the models suggest that both $\lambda > 254$ nm and $\lambda \approx 265$ nm radiation are abundant on the surface of the primeval Earth, the Ritson and Sutherland (2012) pathway should function in both scenarios; however, depending on the nature of the mechanism, the difference between lamp and primeval input may drive significant differences in reaction rate. We note that two reductions are required to produce a single molecule of product.

To explore this question quantitatively, we again conduct a simple numerical experiment to that done in Section 4.2.1. We again represent the two UV light sources by the Rugheimer et al. (2015) surficial emergent spectrum and a Pen-Ray Hg lamp radiating $4.4\times10^4$ erg/cm2/s integrated across the 254 nm line at a distance of 1.9 cm. We again consider fluxes from 200-303 nm, chosen to encompass the 200-300 nm region of photochemical interest while avoiding truncating the wavelength bins of our computed prebiotic UV spectrum. We compute the fraction of incident flux absorbed by $Cu(CN)_3^{2-}$ (I) as a function of wavelength assuming a path length of 1 cm and a $Cu(CN)_3^{2-}$ (I) concentration of 6 mM, chosen to accord with the experimental setup of Ritson and Sutherland (2012). We obtain the spectral molar absorption of $Cu(CN)_3^{2-}$ (I) from Magnani (2015) (see Appendix D). At this concentration and path length, the absorption spectrum is optically thick across most of the wavelength range under consideration, implying the photolysis rate should depend only weakly on the spectral absorbance. Lastly, we represent the QEF of the photoionization-driven mechanism postulated by Ritson and Sutherland (2012) as a step function valued at 1 for $\lambda < \lambda_c$ nm and 0 otherwise, and the QEF of the photoexcitation-driven mechanism postulated by Banerjee et al. (2014) by a Gaussian with amplitude 1, centered at 265 nm, with width $\sigma_c$. We convolved these emission spectra, absorption curve, and QEFs together to determine the HCN photoreduction rate as a function of wavelength. We explored a range of values for $\lambda_c$ and $\sigma_c$. Figures 7a and 7b show the dependence of the integrated photoreduction rate on $\lambda_c$ and $\sigma_c$.

Figure 7a shows the integrated HCN photoreduction rate under exposure to lamp and model prebiotic spectra as a function of $\lambda_c$. The lamp photoreduction rate increases rapidly as $\lambda_c$ increases from 251 to 257 nm, corresponding to the mercury emission line. The photoreduction rate then levels off, reflecting the minimal flux produced outside the emission line. By contrast, the prebiotic photoproduction rate increases monotonically with $\lambda_c$ as more and more of the absorbed flux is usefully exploited. The lamp photoreduction rate is greater than the prebiotic photoreduction rate even at $\lambda_c$=300 nm. This is because the irradiance from the UV lamp at the distance we have modeled here (1.9 cm) is larger in amplitude than the prebiotic flux. A dimmer lamp or larger irradiance distance will lead to lower lamp fluxes and hence photoreduction rates.

Figure 7b shows the integrated photoreduction rate under exposure to lamp and model prebiotic spectra as a function of $\sigma_c$. For values of $\sigma_c < 3.4$ nm, the lamp photoreduction rate is less than the prebiotic photoreduction rate. This reflects the displacement between the 254 nm emission of the lamp and the 265 nm center of the QEF. However, for $\sigma_c > 3.4$ nm, the wings of the QEF are wide enough to capture adequate lamp flux to exceed the photoproduction rate under prebiotic UV input. This is a function of both the wavelength dependence of the action spectrum as well as the greater amplitude of UV flux emanating from the lamp.

Also plotted in both figures is the experimental detection threshold, the photoreduction rate required to generate a detectable quantity of product. The detection threshold is dependent on the laboratory setup and experimental technique used to measure the photoreduction rate, as well as the experimental integration time. The example plotted here is based on an experimental setup which requires require 50 picomols of product to be present in 10 μL samples for an LCMS detection[3]. We assume an integration period of 7 hours for the example presented here, corresponding to the integration period used in Ritson and Sutherland (2012). Under these assumptions, a detectable quantity of product is generated for a wide range of $\lambda_c$ and $\sigma_c$. This suggests experimental measurements of the action spectrum of this photoprocess should be tractable.

---

[3] C. Magnani & A. Bjorkbom, personal communication, 05/15/2015

Figure 8 shows the photoreduction rate calculation for $\lambda_c$=257 nm and $\sigma_c$=3.65 nm. At these values, the photoreduction rates under exposure to the lamp emission spectrum are equal under both mechanisms. However, under prebiotic emission, the Gaussian QEF yields photoreduction rates 69 times higher than the step function QEF. The HCN photoreduction rate can vary by two orders of magnitude depending on whether lamp or modeled prebiotic fluxes are used! This numerical experiment illustrates the importance of characterizing action spectra of prebiotic photoprocesses, to enable extrapolation from laboratory to prebiotic contexts.

**5. Conclusions**

The prebiotic UV environment was exposed to high levels of UV radiation relative to the present day due to lack of UV-shielding $O_2$ and $O_3$. However, environmental constituents likely blocked the shortest-wavelength radiation. Micron-thick films of water can extinct UV flux with $\lambda < 168$ nm, suggesting that even on bodies such as asteroids and comets such short wavelengths are not relevant to prebiotic chemistry. On Earth, attenuation of solar UV flux due to atmospheric carbon dioxide means that UV inputs with $\lambda < 204$ nm was sharply attenuated. However, UV flux with $\lambda > 204$ nm would have been readily available. It is possible that $SO_2$ absorption during epochs of high volcanism may have been comparable or even exceeded that due to $CO_2$;

future work should self-consistently include $SO_2$ absorption at varying levels in estimating atmospheric UV attenuation.

These findings bear lessons for efforts to simulate prebiotic photochemistry in laboratory settings. First, UV sources emitting radiation with $\lambda < 168$ nm, such as 158 nm fluorine lasers, should not be used to simulate aqueous prebiotic chemistry in any situation. For surficial chemistry on the prebiotic earth, sources should emit at wavelengths $\lambda > 204$ nm; sources like ArF eximer lasers which emit at 193 nm, or Hg lamp emission at 184.9 nm, should not be used. Second, broadband UV sources are favored over narrowband sources like UV lamps. Sources able to incorporate the effects of solar atomic lines and molecular absorption, like tunable UV lasers, would provide an even higher-fidelity simulation of the UV environment. Tunable sources will also enable simulation of phenomena such as solar activity, which may influence photochemistry. However, based on proxies for the young Sun such as $\kappa^1$ Ceti and the modern Sun, the variation in UV output of the young Sun may have been comparatively modest, corresponding to a < 1% variation in ribonucleotide photoabsorption rates integrated over 200-300 nm. This suggests variations in UV output driven solar variability are not crucial drivers of prebiotic chemistry. Third, there can be significant differences (1-2 orders of magnitude) between natural UV fluence levels and those generated by UV lamps. The often-higher fluence levels of UV lamps can make the timescales of photochemistry-driven processes more tractable to laboratory study, but such studies should take care when extrapolating from high-fluence regimes to natural conditions. Sensitivity studies characterizing the dependence of reaction rate on fluence level can help with such extrapolations.

Simulations of prebiotic chemistry often rely on the availability of feedstock gases such as HCN and $CH_4$. A key constraint on the availability of these gases is UV photolysis, which would have been higher in the prebiotic era due to higher UV throughput in the atmosphere. Atmospheric $CO_2$ can shield such gases from photolysis, depending on its abundance. For $CH_4$, $N_{CO2} > 7.46\times10^{20}$ cm$^{-2}$ is enough to remove photolytic constraints on $CH_4$ abundance. Under our assumptions of a uniform isothermal atmosphere in equilibrium, achieving this level of shielding at z=50 km requires a $CO_2$ surface pressure of $1.49\times10^{-2}$ bar, with lower $CO_2$ levels being required the deeper one goes into the atmosphere. $CO_2$ surface pressures of $> 2.77\times10^{-4}$ bar will remove photolytic constraints on tropospheric (z < 17 km) $CH_4$. The more photoactive HCN requires more shielding: $N_{CO2} > 2.35\times10^{23}$ cm$^{-2}$ is required to lift photolytic constraints up to an altitude of 50 km, corresponding to a staggering 4.75 bars of $CO_2$ required at the surface. If HCN is sequestered in the troposphere, then $8.76\times10^{-2}$ bar of $CO_2$ is adequate to remove the photolytic constraint.

The works of Ritson and Sutherland (2012) and Powner et al. (2009) towards prebiotic synthesis of RNA provide case studies into the relevance of accurately reproducing the UV environment. The Ritson and Sutherland (2012) pathway for the controllable synthesis of simple sugars uses UV light to reduce HCN using tricyanocuprate (I) as an electron donor. Two theoretical mechanisms have been suggested whereby this is accomplished. We have demonstrated that under both mechanisms, narrowband 254 nm mercury lamp radiation is an acceptable proxy for prebiotic UV flux with the caveat that, depending on the underlying mechanism, reaction rates measured under narrowband lamp radiation may conceivably be different from those measured under irradiation corresponding to the prebiotic environment by multiple orders of magnitude.

The Powner et al. (2009) pathway for the synthesis of activated ribonucleotides uses UV light in two steps. The most crucial of these is the amplification of ribocytidine and ribouridine relative to other products of the synthesis reaction due to differential UV photolysis. The wavelength dependence of this relative photostability advantage is not known for this set of molecules. It is plausible that photolysis rates will vary as a function of molecule and wavelength. We have demonstrated that it is possible for there to exist photolysis quantum efficiency curves such that the Powner et al. (2009) process proceeds in laboratory settings but not in a radiation environment corresponding to the surface of the young Earth. Further experimental work, e.g. replicating the experiment under broadband or tunable laser UV input, is required to fully validate this pathway.

In general, our work suggests the importance of characterizing the action spectra (wavelength dependence) of UV-sensitive chemical pathways thought relevant to prebiotic chemistry. Determining these action spectra enables the extension of laboratory studies to the natural prebiotic environment. Further, these action spectra can be used to determine the viability of proposed chemical pathways on alien planetary environments, e.g. exoplanets. Characterizing the action spectra of prebiotic pathways is crucial to understanding whether mechanisms hypothesized to have lead to abiogenesis on Earth could also operate on other planets.

## Acknowledgements

We thank C. Magnani for many discussions and sharing his measurement of the tricyanocuprate absorption spectrum with us. We thank S. Rugheimer for sharing her prebiotic atmosphere models and insight with us. We thank M. Powner for sharing insights on the photochemistry underlying the pathways in his work with us, as well as further discussions and his feedback on our manuscript. We additionally thank I. Ribas for kindly sharing his spectrum of $\kappa^1$ Ceti and his insight upon request, I. Cnossen for sharing her young Sun models, C. Y. Wu for sharing his methane cross-section data, V. Vuitton for sharing her compilation of HCN cross-sections, and E. Schwieterman for help with absorption spectra of atmospheric absorbers. We thank D. Zubarev, D. Ritson, K. Zahnle, R. Wordsworth, F. Tian, J. Delano, S. Harman, A. Bjorkbom, A. Glenday, and A. Fahrenbach for instructive discussions. We thank L. Schaefer and an anonymous reviewer for comments that substantially improved this manuscript.

This research has made use of NASA’s Astrophysics Data System Bibliographic Services, and the MPI-Mainz UV-VIS Spectral Atlas of Gaseous Molecules.

S. R. and D. D. S. gratefully acknowledge support from the Simons Foundation, grant no. 290360.

## Author Disclosure Statement

The authors declare no competing financial interests.

**Tables**

Table 1: $CO_2$ levels required to remove constraint of photolysis on buildup of $CH_4$ for varying values of $z_0$.

| | $z_0$=50 km | $z_0$=17 km | $z_0$=1 km |
|---|---|---|---|
| $N_{CO2}$ ($cm^{-2}$) | $7.46\times10^{20}$ | $7.41\times10^{20}$ | $6.63\times10^{20}$ |
| $P_{CO2}(z=z_0)$ (bar) | $3.60\times10^{-5}$ | $3.57\times10^{-5}$ | $3.19\times10^{-5}$ |
| $P_{CO2}(z=0)$ (bar) | $1.49\times10^{-2}$ | $2.77\times10^{-4}$ | $3.60\times10^{-5}$ |

Table 2: $CO_2$ levels required to remove constraint of photolysis on buildup of HCN for varying values of $z_0$.

| | $z_0$=50 km | $z_0$=17 km | $z_0$=1 km |
|---|---|---|---|
| $N_{CO2}$ ($cm^{-2}$) | $2.38\times10^{23}$ | $2.35\times10^{23}$ | $2.05\times10^{20}$ |
| $P_{CO2}(z=z_0)$ (bar) | $1.15\times10^{-2}$ | $1.13\times10^{-2}$ | $9.89\times10^{-3}$ |
| $P_{CO2}(z=0)$ (bar) | 4.75 | $8.76\times10^{-2}$ | $1.12\times10^{-2}$ |

## Figures

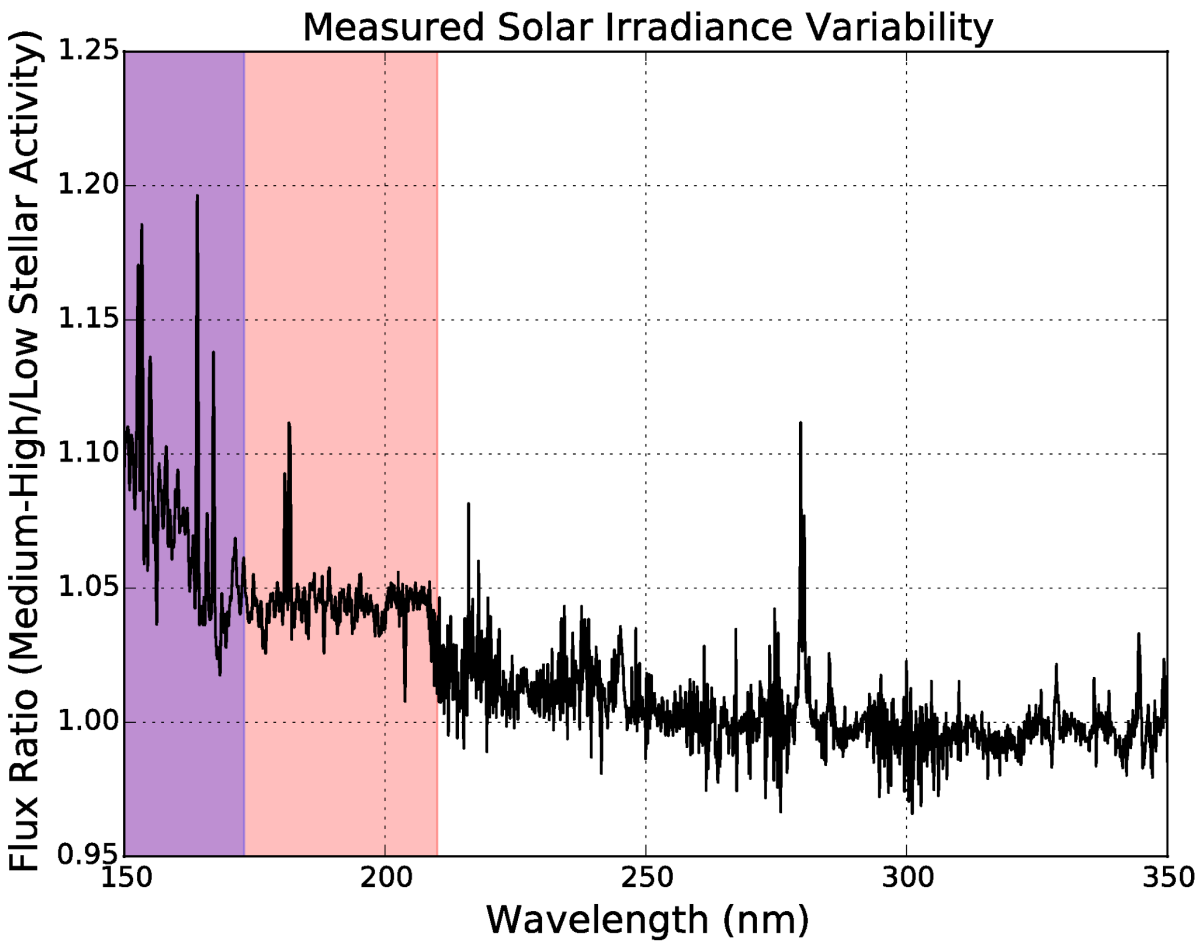

(a) Ratio of composite solar reference spectra from Thuillier et al. (2004) corresponding to the ATLAS 1 (moderately-high activity) and ATLAS 3 (low activity) missions.

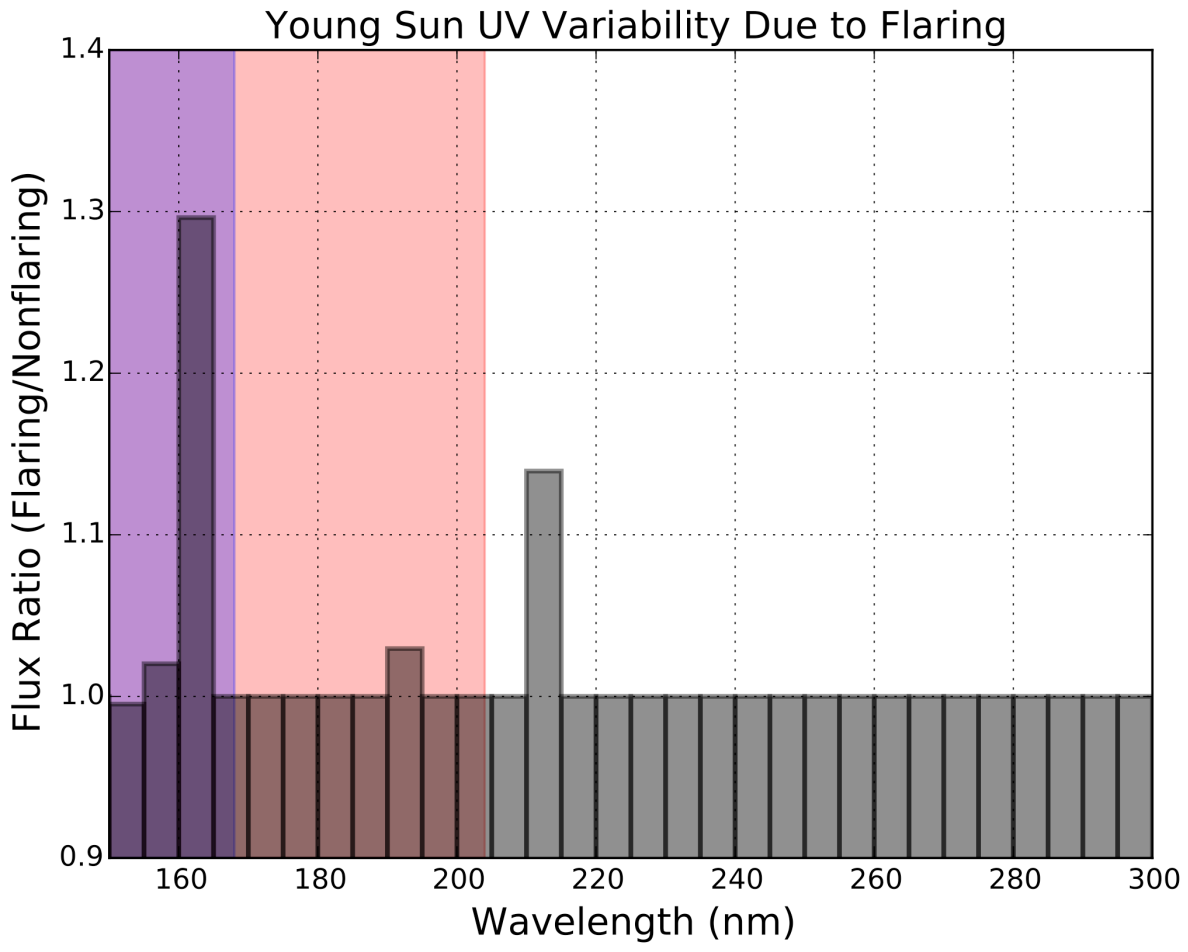

(b) Ratio of model spectra of flaring and non-flaring young Sun derived by Cnossen et al. (2007) for $\lambda$=150 − 300 nm.

Fig. 1.: Two estimates for variability in the emission spectrum of the young

Sun. Shaded in red is the region of the spectrum shielded by the 3.9 Ga atmosphere. Shaded in purple is the region of the spectrum shielded by a 1μm layer of water. The impact of flux variations in these regions will be damped by atmospheric and aqueous attenuation.

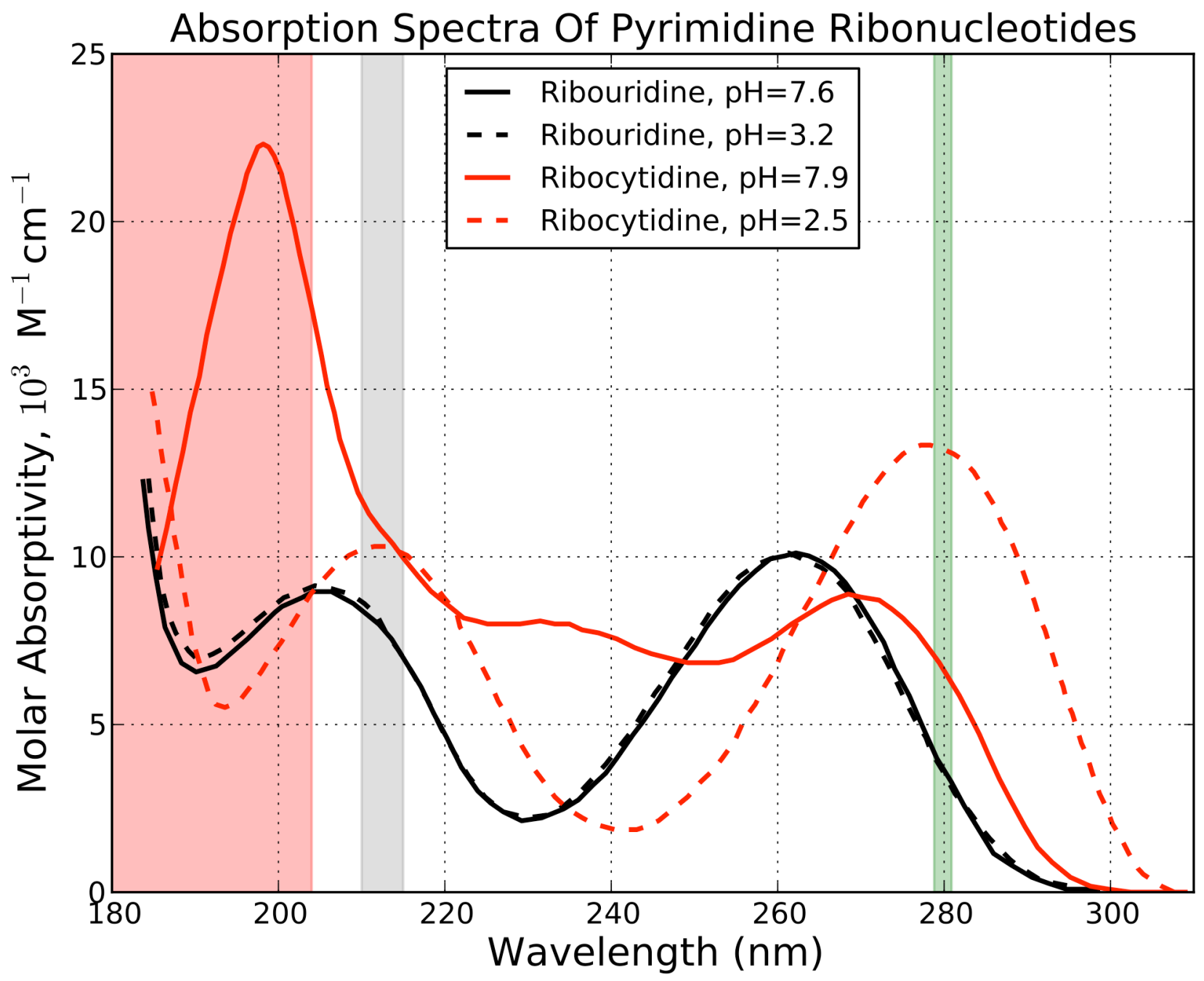

Fig. 2.: Absorption spectrum of pyrimidine ribonucleotides at different pH's, taken from Voet et al. (1963). Shaded in red is the region of the spectrum shielded by the prebiotic atmosphere. Shaded in grey is the region of the spectrum corresponding to the 210-215 nm variability feature due to flaring from the work of Cnossen et al. (2007). Shaded in green is the region of the spectrum corresponding to the Mg II k, h line complex identified from the data of Thuillier et al. (2004).

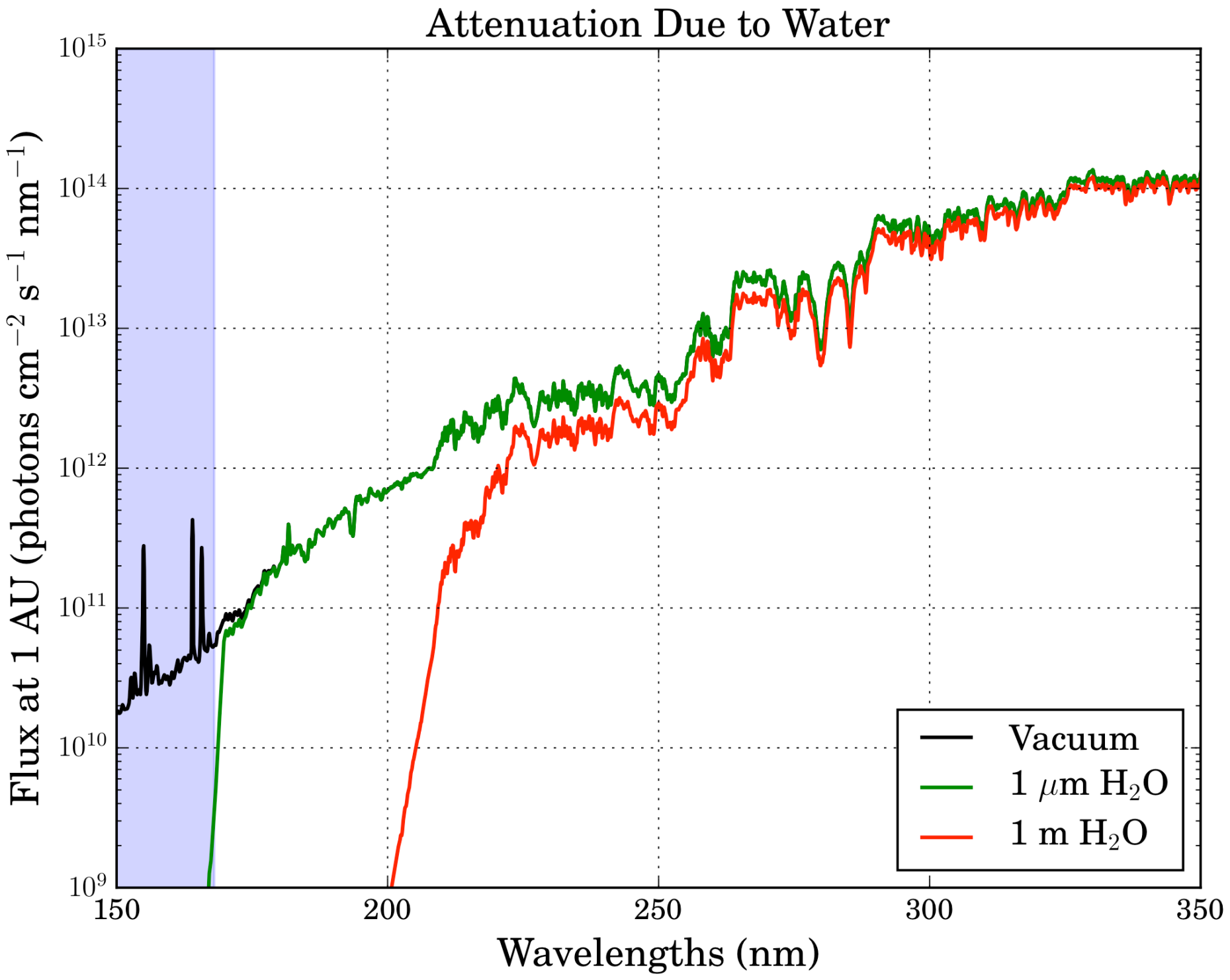

Fig. 3.: Attenuation of UV spectrum of 3.9 Ga sun through water layers of varying thickness. The spectral region shielded by absorption from a 1µm layer of water is shaded in blue.

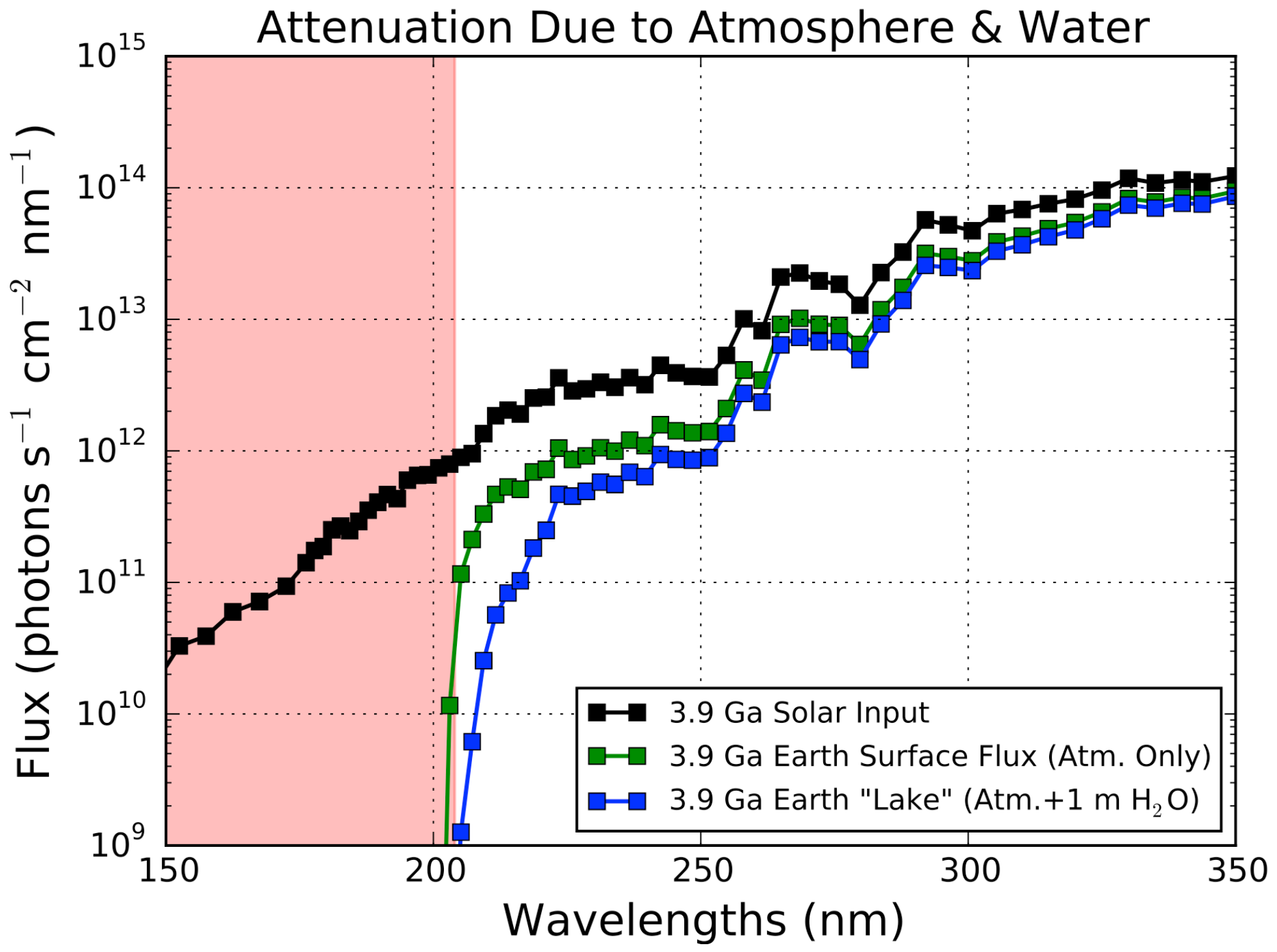

Fig. 4.: Attenuation of the 3.9 Ga solar UV spectrum due to the atmosphere modeled by Rugheimer et al. (2015), corresponding to the surface UV flux. Also shown is the surface flux attenuated by 1 m of water, corresponding to a surficial lake. The spectral region shielded by atmospheric absorption (primarily $CO_2$) is shaded in red.

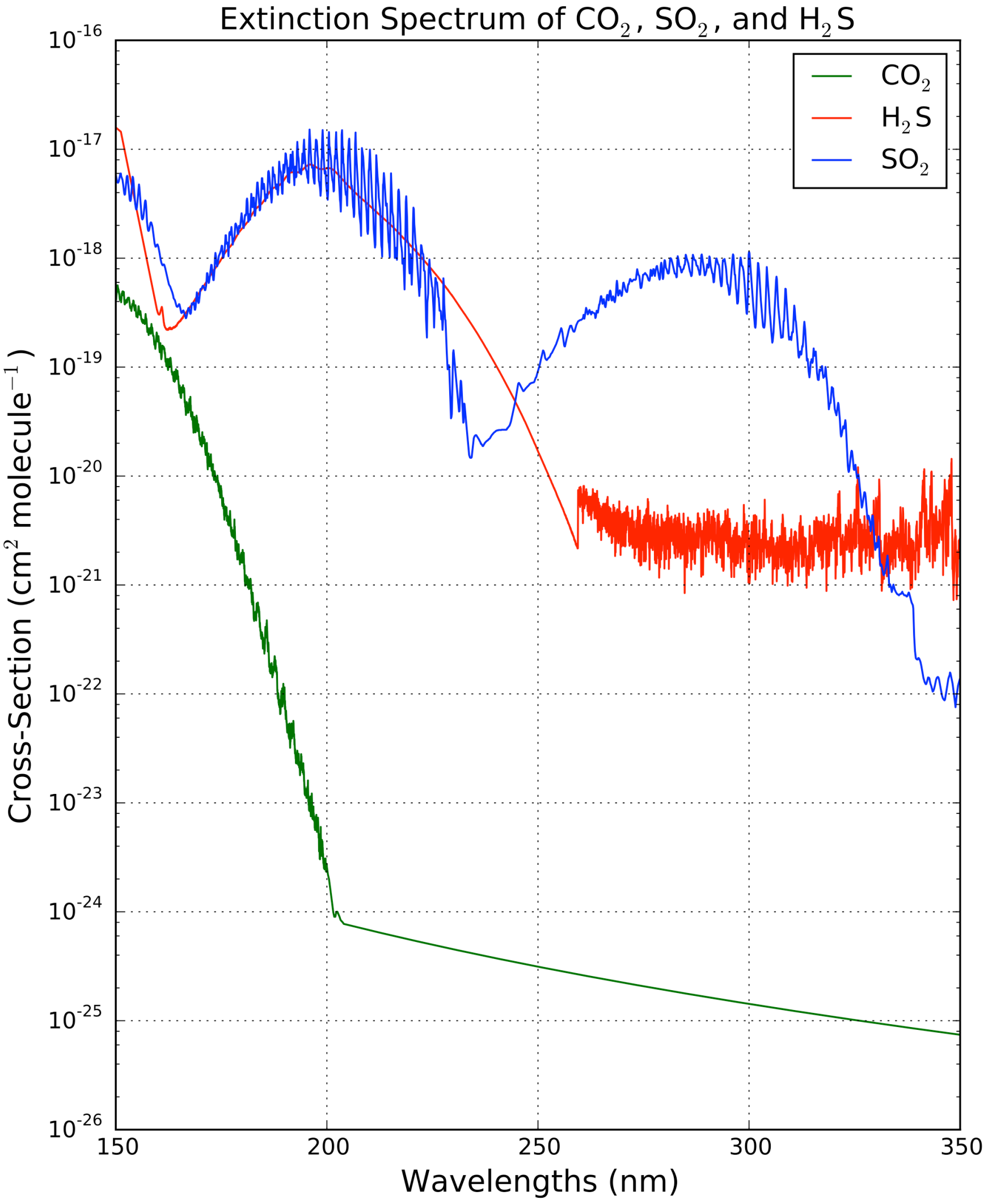

Fig. 5.: UV extinction cross-sections of $SO_2$, $H_2S$, and $CO_2$. Appendix E describes the sources of these cross-sections.

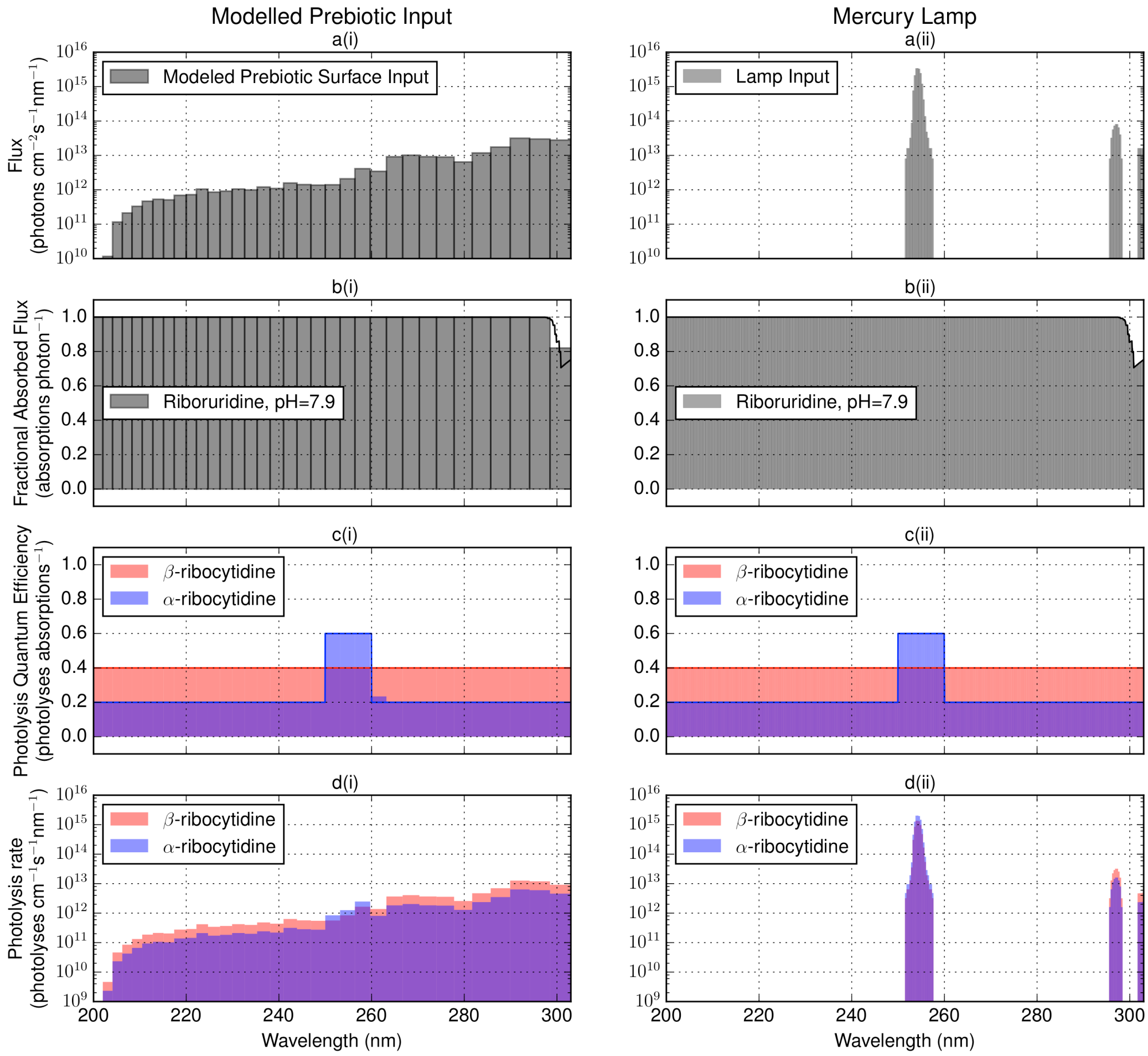

Fig 6.: (a) Incident flux, (b) fraction of incident flux absorbed, (c) assumed QEF of photolysis for β (biogenic) and α (nonbiogenic) ribocytidine. Panel (d) presents photolysis rate for each stereoisomer under irradiation by (i) prebiotic flux and (ii) mercury lamp flux, formed by convolving panels (a), (b) and (c).

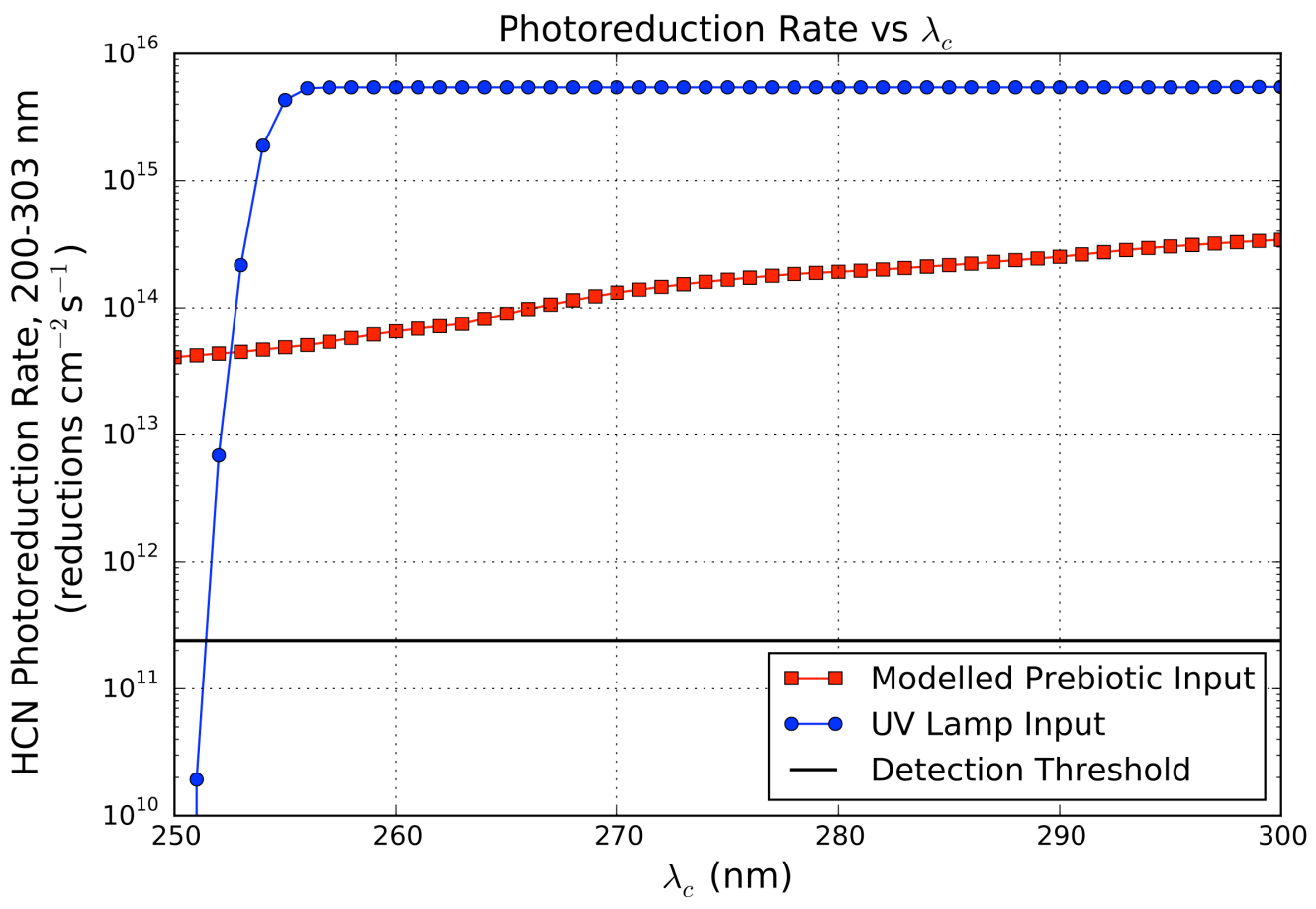

(a) Integrated HCN photoreduction rate (200-303 nm) under lamp and modeled prebiotic UV input as a function of $\lambda_c$ (the threshold value for the assumed step function QEF)

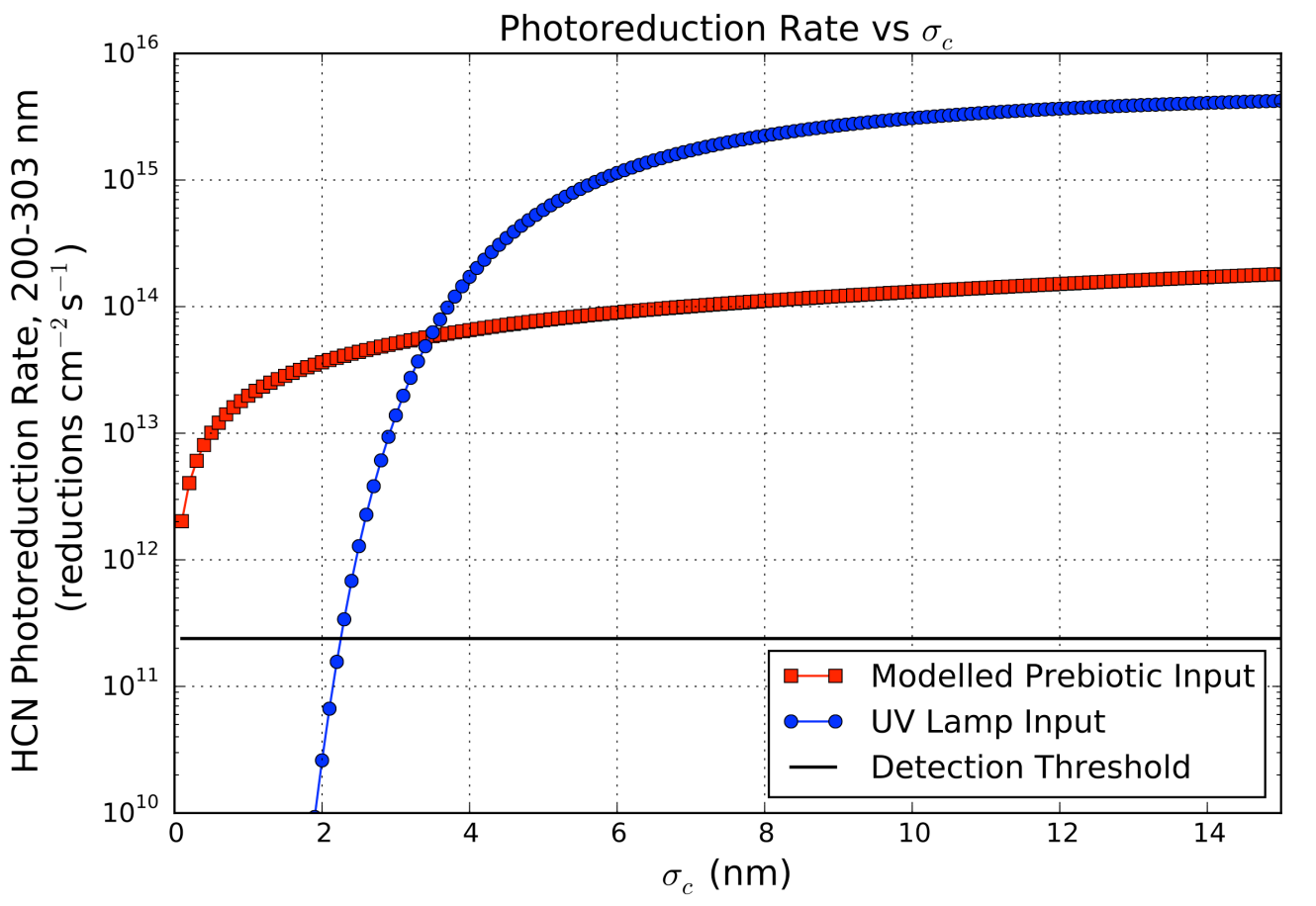

(b) Integrated HCN photoreduction rate (200-303 nm) under lamp and modeled prebiotic UV input as a function of $\sigma_c$ (the standard deviation of the assumed Gaussian profile for the reaction QEF).

Fig. 7.: Dependence of HCN photoreduction rate on $\lambda_c$ and $\sigma_c$. Also plotted is the minimum photoreduction rate required to generate a detectable quantity of product assuming 7-hour integration, 10-μL sample, and 50-picomol detection threshold.

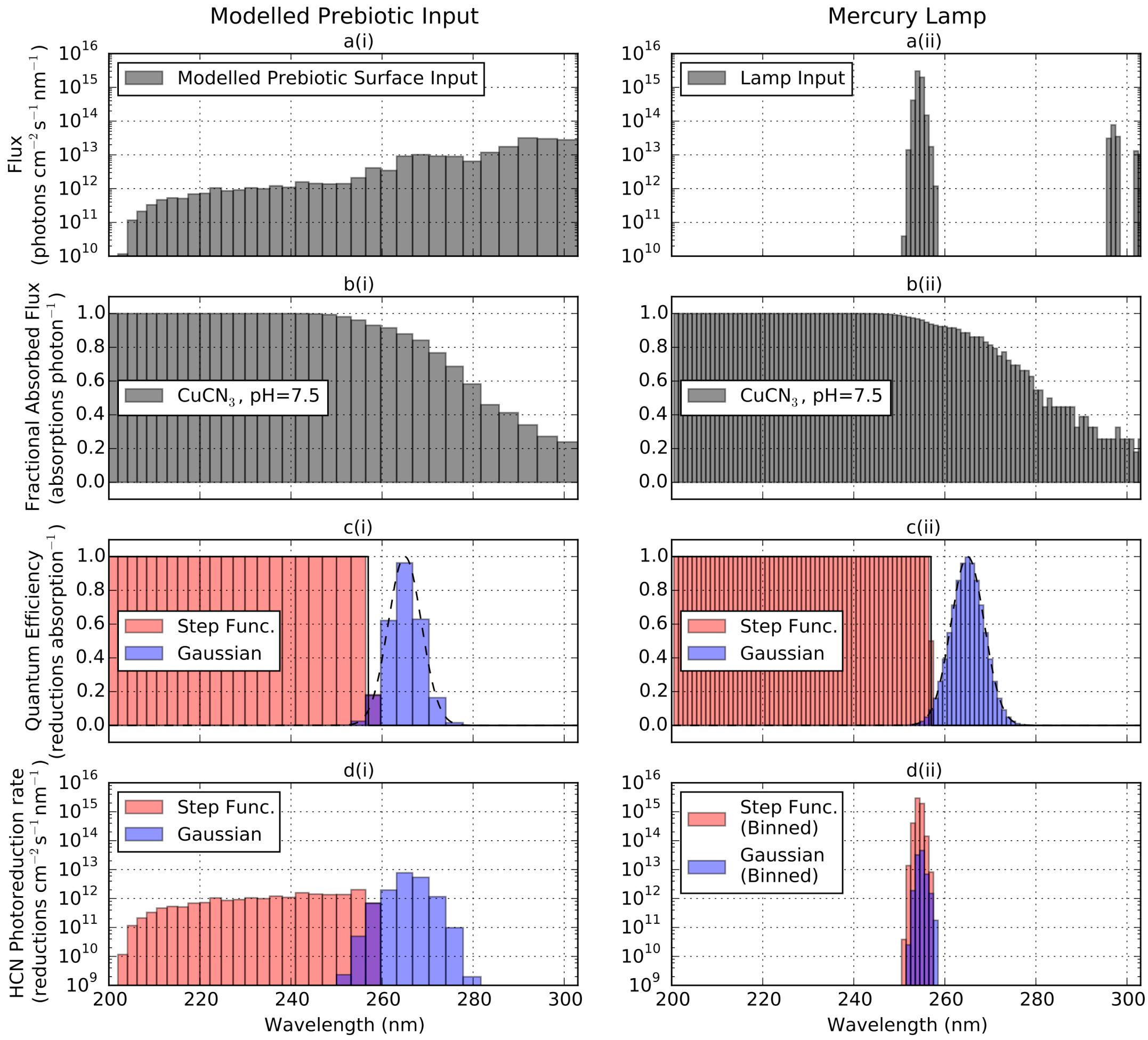

Fig. 8.: (a) Incident flux, (b) fraction of incident flux absorbed, (c) assumed Gaussian and step function QEFs of HCN photoreduction by tricyanocuprate. Panel (d) presents photoreduction rate for the two assumed QEFs under irradiation by (i) prebiotic flux and (ii) mercury lamp flux, formed by convolving panels (a), (b) and (c).

## Appendix A. Constraints on the Era of Abiogenesis

This Appendix presents the literature review underlying our decision to choose 3.9 Ga as the time period relevant to abiogenesis.

Based on evidence such as the discovery of fossilized microorganisms (microfossils) (Javaux et al. 2010) and microbially induced sedimentary structures (MISS) generated by microbial mats (Noffke et al. 2006), life was established by 3.2 Ga. Reports of putative stromatolites[4] (Buick 2007; Hofmann et al. 1999), MISS (Noffke et al. 2013), and potentially biogenic carbon and sulfur fractionation (Buick 2007) are suggestive of life being established by 3.5 Ga. Carbon fractionation arguments have been used to argue for life as early as 3.7 Ga, but this is still debated (Ohtomo et al. 2013; Buick 2007). For purposes of this work, therefore, we constrain the era of abiogenesis to have been earlier than 3.5 Ga.

Upper bounds on the era of abiogenesis suffer from even more paucity of information. Surface liquid water existed as early as 4.3 Ga (Mojzsis et al. 2001; Catling and Kasting 2007). Another constraint on the era of abiogenesis may come from the Late Heavy Bombardment (LHB). From 4.1-3.8 Ga, the Earth experienced a high flux of extraterrestrial impactors (Deamer 2007), peaking sharply 3.9 Ga (Chapman et al. 2007). Maher and Stevenson (1988) argue that bombardment would have sterilized the planetary surface and would have heated the oceans, potentially to uninhabitable levels. Sleep et al. (1989) go further, arguing that LHB impacts could vaporize the oceans. However, Abramov and Mojzsis (2009) incorporate better cratering records and new Solar System dynamical and terrestrial lithospheric models to conclude that the 3.9 Ga peak of the LHB may not have entirely sterilized life. We cannot therefore state with certainty that the LHB sterilized the Earth. Consequently, abiogenesis may have occurred as early as 4.3 Ga.

[4] Laminated sedimentary structures that trace microbial activity.

Hence, the range of ages potentially relevant to prebiotic chemistry is 3.5-4.3 Ga. We choose 3.9 Ga as it lies squarely in the middle of this range and postdates the LHB, hence avoiding planetary sterilization concerns.

**Appendix B. The 3.9 Ga Terrestrial Environment**

This Appendix summarizes available constraints on the prebiotic atmosphere at 3.9 Ga.

Evidence from zircon samples demonstrates the existence of surface liquid water starting 4.3 Ga ago (Mojzsis et al. 2001; Catling and Kasting 2007). This means that the surface temperature had to be above the freezing point of water. This is a key constraint as the young Sun was as much as 30% fainter compared to the present day. In order to maintain surface liquid water an enhanced greenhouse effect due to higher levels of $CO_2$ is typically invoked (Kasting 1993; Kasting 2014).

The atmosphere in the prebiotic era was likely dominated by volcanic outgassing from high-temperature magmas. The Earth's interior had differentiated by 4.4 Ga, meaning that the composition of the mantle was set by this date (Catling and Kasting 2007). Delano (2001) study Cr and V abundance in ancient volcanic rocks to show that the redox state of volcanic rocks and hence the mantle has not changed since 3.6 Ga, and likely since 3.9 Ga[5]. Since the redox state controls speciation of H-C-O-S elements, this fact implies that primordial volcanic outgassing of H-C-O-S elements from T>1300K magma was dominated by $H_2O$, $CO_2$, and $SO_2$, with ≤ 1% contributions of $H_2$ and CO. Trail et al. (2011) extend this constraint by studying incorporation of Ce into ancient zircons. Their analysis suggests that the redox state of the mantle has been unchanged to within measurement error since 4.3 Ga, and conclude that volcanic outgassing was dominated by $H_2O$, $CO_2$, $SO_2$, and $N_2$, consistent with Delano (2001). Zolotov and Shock (2000) give the relative abundances of $H_2O$, $CO_2$, and $SO_2$ in a quartz-fayalite-magnetite (QFM) buffer, which approximates the terrestrial mantle, as a function of temperature. For T>1000ºC, the abundances of $H_2O$, $CO_2$, and $SO_2$ are roughly independent of temperature; for T=1200ºC, the mole fractions of $H_2O$, $CO_2$, and $SO_2$ are 0.54, 0.31, and 0.15, respectively. It is generally assumed that gases are given off by magmas in the same proportions as they were dissolved[6]. Despite its high production rate, $SO_2$ is not expected to be a major component of the atmosphere due to its tendency to photolyze or be oxidized into rock. However, during periods of high volcanism it is plausible for $SO_2$ to build up to the level of hundreds of parts per million (ppm) (Kaltenegger and Sasselov 2010). Therefore, particularly on the more active primordial Earth, $SO_2$ may have at least transiently been an important component of the planetary atmosphere.

---

[5] The paucity of high-quality samples makes it challenging to draw fully robust conclusions for >3.6 Ga, though what data exists is suggestive.

[6] J. Delano, private communication 7/9/2014

Molecular nitrogen ($N_2$) is another key component of the terrestrial atmosphere, and may have been delivered by impacting planetismals (Kasting 2014). Factors affecting nitrogen abundance in the prebiotic era include production rates from planetismals, which would have been higher on the young Earth. However, atmospheric shock heating of nitrogen in a nonreducing (e.g. $N_2/CO_2$) atmosphere can also convert $N_2$ to NO, which can be geochemically fixed and removed from the atmosphere (Summers et al. 2012). Goldblatt et al. (2009) argue that the nitrogen reservoirs in the mantle are subducted, not primordial, and that crustal nitrogen accumulated with the continents; based on these and the utility of high $N_2$ levels for greenhouse warming to solve the Faint Young Sun paradox, they argue that primordial $N_2$ levels were as much as 2-3 times the present level of 0.79 bar. However, Marty et al. (2013) use $N_2$/Ar fluid inclusions in 3.5 Ga quartz crystals to demonstrate that $N_2$ pressures and isotopic composition at this time were, similar to present-day levels (0.5-1.1 bar).

In summary, existing evidence is suggestive of an $N_2/CO_2$-dominated atmosphere with sufficient concentration of greenhouse gases (e.g. $CO_2$, $CH_4$) to support liquid water on the surface. During epochs of high volcanism, volcanogenic gases (especially $SO_2$) may also have formed an important constituent of the atmosphere. If the young Earth were warm, water vapor would also be a significant atmospheric constituent.

**Appendix C. Derivation of Photolytic Constraints on Feedstock Gas Buildup**

In this section, we derive the equilibrium surface partial pressure of a feedstock gas *X*, assuming the sole sink to be photolysis. As discussed in Section 3.4, we assume an isothermal atmosphere of ideal gases in hydrostatic equilibrium dominated by 0.9 bar $N_2$ and 0.1 bar $CO_2$, with total surface pressure 1 bar. We assume the mixing ratio of the feedstock gas under consideration to remain constant until a height $z_0$, whereupon it goes to zero. Lastly, we assume the feedstock gas population to be optically thin. We take our top-of-atmosphere input $\phi_0(\lambda)$ to be the UV spectrum of the 3.9 Ga sun computed from the models of Claire et al (2012) spanning a wavelength range of 1-300 nm at 0.1 nm resolution. We obtain our $CO_2$ cross-sections $\sigma_{CO2}(\lambda)$ from Huestis and Berkowitz (2010), who compile $CO_2$ cross-sections at 300 K from 0.12-201.6 nm.

Let the rate of supply of methane to the atmosphere to be *S*. Let the rate of removal of *X* from the atmosphere due to photolysis be *B*. Then *B* satisfies

$$d^2B = d\lambda dV n_X(r)\sigma_X(\lambda)\phi(\lambda)\cos(\Phi),$$

where $n_X(r)$ is the number density of $X$ as a function of $r$, $\sigma_X(\lambda)$ is the cross-section of $X$ as a function of wavelength, $\phi(\lambda)$ is the emergent flux irradiating $X$, $r$ is the distance from the planet center, and $\Phi$ is the zenith angle of the irradiating flux. Then,

$$B = \int d\lambda \int dV n_X(r)\sigma_X(\lambda)\phi(\lambda)\cos(\Phi)$$

$$B = \left[\int d\lambda \sigma_X(\lambda)\phi(\lambda)\right]\left[\int_0^{2\pi} d\theta \int_0^{\pi/2} d\Phi \sin(\Phi)\cos(\Phi) \int_{R_{Earth}}^{R_{Earth}+z_0} r^2 dr n_X(r)\right]$$

Where $R_{Earth}$ is the radius of the Earth. $z_0/R_{Earth} << 1$, so changing variables to $z=r-R_{Earth}$, we can write

$$B = \left[\int d\lambda \sigma_X(\lambda)\phi(\lambda)\right]\left[2\pi \times \frac{1}{2}\right]\left[\int_{R_{Earth}}^{R_{Earth}+z0} r^2 dr n_X(r)\right]$$

$$B \approx \pi R_{Earth}{}^2 \times \left[\int d\lambda \sigma_X(\lambda)\phi(\lambda)\right]\left[\int_0^{z0} dz n_X(z)\right]$$

Under our assumptions of an isothermal atmosphere of ideal gases in hydrostatic equilibrium, we can compute:

$$\int_0^{z_0} dz n_X(z) = \int_0^{z_0} dz \frac{P_X(z=0)}{kT}\exp(-\frac{z}{H})$$

$$\int_0^{z_0} dz n_X(z) = \frac{P_X(z=0)}{kT} \times H(1-\exp(-\frac{z_0}{H}))$$

Where the scale height $H=(kT)/(\mu g)$, $\mu$ is the mean molecular mass of the atmosphere, $g$ is the acceleration due to gravity, $T$ is the temperature of the atmosphere. For a 0.1 bar $CO_2$, 0.9 bar $N_2$ atmosphere, mu=29.6 amu. We take $T=290$ K, the surface temperature computed in the model of Rugheimer et al (2015). Then $H=8.3$ km, comparable to $H=8.5$ km for the modern Earth.

We have labelled the input flux at the top of the atmosphere (TOA) to be $\phi_0(\lambda)$. Then

$$\phi(\lambda) = \phi_0(\lambda)\exp(-\tau_{CO2}(\lambda))$$
$$\phi(\lambda) = \phi_0(\lambda)\exp(-N_{CO2}\sigma_{CO2}(\lambda))$$

Where $N_{CO2}$ is the column density of $CO_2$ shielding *X*. We can then compute the quantity

$$b_X = \int_{\lambda_{\min}}^{\lambda_{\max}} d\lambda \sigma_X(\lambda)\phi(\lambda)$$

Where $\lambda_{min}$ and $\lambda_{max}$ correspond to the limits of the available absorption data for *X*. As an example, for $N_{CO2}$=1.4×10$^{21}$ cm$^{-2}$, $b_{CH4}$=3.1×10$^{-27}$ s$^{-1}$.

Combining these relations by equating the source and sinks of *X*, we may reason:

$$S = B$$
$$S = \pi R_{Earth}^2 b\left[\frac{P_X(z=0)}{kT} \times H\left(1-\exp(\frac{-z_0}{H})\right)\right]$$
$$\Rightarrow P_X(z=0) = \frac{kTS}{\pi R_{Earth}^2 bH\left(1-\exp(\frac{-z_0}{H})\right)}$$

**Appendix D. Collection of $Cu(CN)_3^{2-}$ Absorption Spectrum**

The absorption spectrum of tricyanocuprate (I) used in this paper was taken from Magnani (2015), an undergraduate thesis. While the content of this thesis is currently being prepared for publication, this material is not yet peer-reviewed and publicly available. Hence, with permission of the author, we briefly excerpt here a description of the techniques used to collect this spectrum. We emphasize that credit for this work must go to Magnani (2015) and their forthcoming paper.

The absorption spectra in this study were collected using a Starna quartz cuvette (1 cm cube) mounted in a UV-Vis spectrometer (Ultraspec 3100 Pro, Amersham Biosciences). First, an absorption spectrum was taken of a blank consisting of 2.5 mL of pure deionized water. Then, a $Cu(CN)_3^{2-}$ solution was prepared by combining a stock solution of 0.2 mM CuCN with a 1.1 equivalent of $CN^-$. A spectrum of this solution was taken using the UV-Vis spectrometer and quartz cuvette. Additional spectra were taken of the complex as 0.1 equivalents of cyanide were added, up until a total of 5 equivalents of the original 0.2 mM CuCN solution. Additionally, a spectrum of a 0.89 mM $CN^-$ solution formed by dissolving KCN in deionized degassed water was collected, allowing the determination of the absorption spectrum of pure $CN^-$. The same cuvette was used for all absorption experiment for consistency. Finally, the blank was differenced from the other spectra to subtract out background absorption due to water, the cuvette, and oxygen in the air.

The titration of CuCN by CN demonstrated that as CN concentration increased, the solution spectrum increasingly resembled the spectrum of pure $CN^-$. This indicates that the additional CN was being partitioned preferentially into solution as opposed to into the tricyanocuprate complex. Therefore, we chose to use the spectrum corresponding to the solution formed by combining of 0.2 mM CuCN with just 1.1 equivalent of $CN^-$ to represent the spectrum of the tricyanocuprate (I) complex; we term this the “base solution”. The tricyanocuprate species dominates the spectrum because the dicyanocuprate species is not photoactive and the tetracyanocuprate species is not thermodynamically favored at neutral pH (Horváth et al. 1984).

The base solution was prepared by combining a stock solution of 0.2 mM CuCN with a 1.1 equivalent of $CN^-$, resulting in a solution containing 0.2 mM $Cu^+$ and 0.42 mM $CN^-$. Assuming all the $CN^-$ goes into the tricyanocuprate complex, this corresponds to a 0.14 mM solution of tricyanocuprate (I). However, the partitioning between the $CN^-$ in solution versus in the tricyanocuprate complex is not known. Hence, 0.14 mM represents an upper limit on the tricyanocuprate (I) concentration.

We computed the molar absorptivities via the absorbance formula $A/(d \times C)=e$, where A is the absorbance, e is the molar absorptivity ($M^{-1}cm^{-1}$), d is the path length in cm, and C is the concentration of the solution in M. $A=\log_{10}(I_0/I)$, where $I_0$ is the intensity incident on the cuvette and I is the transmitted intensity. A is reported by the UV-Vis spectrometer, the cuvette dimensions set d=1 cm and we took C=0.14 mM. As 0.14 mM is an upper bound for C, this means that the values we compute for e are lower bounds. Ritson and Sutherland (2012) report they formed their tricyanocuprate complex by adding 30 mg KCN (65.12g/mol) and 2 mg CuCN (89.56 g/mol) to 2.2 mL of water and titrating with an unspecified amount of 1M HCl to reach neutral pH. The amount of HCl solution added cannot exceed 1.3 mL as the total volume of the cuvette used in their experiment was 3.5 mL. This corresponds to a total of $2.23\times10^{-5}$ mol $Cu^{+}$ and $4\times10^{-4}$ mol $CN^{-}$. Assuming all the Cu goes into the complex, which seems reasonable given that [Cu]<<3[CN], this corresponds to a tricyanocuprate concentration of $(2.23\times10^{-5}\text{mol})/(2.2-3.5\text{mL})$=6-12 mM. As the spectrum is saturated at the Ritson and Sutherland (2012) concentration of 6 mM (see Figure 8) using even the lower of these values and our lower bounds for e the absorption spectrum, the use of our lower bounds for e has minimal impact on our results.

## Appendix E. Extinction Cross-Sections

This Appendix specifies the sources of the extinction cross-sections for the gases used in Figure 5 (Section 3.3.2). We used laboratory measurements of cross-sections where available, and assumed extinction was due to Rayleigh scattering otherwise. Unless otherwise stated, all measurements were collected near room temperature (295-298 K) and 1 bar of atmospheric pressure, and the digitized data files of the empirical measurements are collected from the MPI-Mainz UV/Vis Spectral Atlas.

### *E.1. $CO_2$*

We take empirically measured cross-sections shortward of 201.6 nm from Huestis and Berkowitz (2010). Huestis and Berkowitz (2010) review existing measurements of extinction cross-sections for $CO_2$, and aggregate the most reliable ones into a single spectrum (< 1 nm resolution). They test their composite spectrum with an electron-sum rule, and find it to agree with the theoretical expectation to 0.33%. From 201.75-300 nm, the measurements of Shemansky (1972) provide coverage. However, the resolution of these data ranges from 0.25 nm from 201.75-203.75 nm, to 12-25 nm from 210-300 nm. Further, Shemansky (1972) finds that essentially all extinction at wavelengths longer than 203.5 nm is due to Rayleigh scattering (Ityaksov et al. 2008 derive similar results). Therefore, we adopt the measurements of Shemansky (1972) from 201.75-203.75 nm, and take Rayleigh scattering to describe CO2 extinction at longer wavelengths.

We compute the Rayleigh scattering cross-section of CO2 using the formalism of Vardavas and Carver (1984): $\sigma=4.577\times10^{-21}\times KCF\times[A(1 + B/\lambda^2)]/\lambda^4$, where $\lambda$ is in μm and KCF is the King correction factor, and where $KCF=(6 + 3\delta)/(6 - 7\delta)$, where $\delta$ is the depolarization factor. This approach accounts for the wavelength dependence of the index of refraction but assumes a constant depolarization factor. We take the values of the coefficients A and B from Keady and Kilcrease (2000), and the depolarization factor of $\delta=0.0774$ from Shemansky (1972).

***E.2. $SO_2$***

From 106.1-403.7 nm, we take the cross-sections for $SO_2$ extinction from the compendium of $SO_2$ cross-sections of Manatt and Lane (1993) (0.1 nm resolution). Manatt and Lane (1993) evaluate extant UV cross-sections for $SO_2$ extinction, and aggregate the most reliable into a single compendium covering this wavelength range at 293 ± 10 K.

### *E.3. $H_2S$*

We take empirically measured cross-sections of $H_2S$ shortward of 159.465 nm from the dipole (e,e) spectroscopy measurements of Feng et al. (1999) (< 10 nm resolution). From 159.465-259.460 nm, we take the cross-sections for $H_2S$ extinction from the gas cell absorption measurements of Wu and Chen (1998) (0.06 nm resolution), as recommended by Sander et al. (2011). From 259.460-370.007 nm, we take the gas cell absorption measurements of Grosch et al. (2015) (0.018 nm resolution). Many of the cross-sections reported in the Grosch et al. (2015) dataset are negative, corresponding to an increase in flux from traversing a gas-filled cell. These cross-sections are deemed unphysical and removed from the dataset. Further, the Grosch et al. (2015) dataset shows a great deal of high-resolution structure that is not relevant to our analysis. We use an 11-point mean boxcar filter to smooth the data we take from the Grosch et al. (2015) datatset.